\documentclass[12pt,preprint]{emulateapj}

\usepackage{natbibspacing,url,txfonts,natbib,longtable,xcolor}

\newcommand{\atone}{{\it apogee\_target1} }
\newcommand{\attwo}{{\it apogee\_target2} }
\newcommand{\teff}{$T_{\rm eff}$ }

\shorttitle{APOGEE Target Selection}

\shortauthors{Zasowski et al.}

\begin{document}

\title{Target Selection for the Apache Point Observatory Galactic Evolution Experiment (APOGEE)}

\author{
G.~Zasowski\altaffilmark{1,2,3,4},
Jennifer~A.~Johnson\altaffilmark{2,3}, P.~M.~Frinchaboy\altaffilmark{5}, S.~R.~Majewski\altaffilmark{4}, D.~L.~Nidever\altaffilmark{6}, 
H.~J.~Rocha Pinto\altaffilmark{7,8}, L.~Girardi\altaffilmark{7,9}, B.~Andrews\altaffilmark{2},
S.~D.~Chojnowski\altaffilmark{4}, K.~M.~Cudworth\altaffilmark{10},
K.~Jackson\altaffilmark{5}, 
J.~Munn\altaffilmark{11}, M.~F.~Skrutskie\altaffilmark{4}, 
R.~L.~Beaton\altaffilmark{4}, C.~H.~Blake\altaffilmark{12}, K.~Covey\altaffilmark{13}, R.~Deshpande\altaffilmark{14,15}, C.~Epstein\altaffilmark{2}, 
D.~Fabbian\altaffilmark{16,17}, S.~W.~Fleming\altaffilmark{14,15}, A.~Garcia~Hernandez\altaffilmark{16,17},
A.~Herrero\altaffilmark{16,17}, S.~Mahadevan\altaffilmark{14,15}, Sz.~M{\'e}sz{\'a}ros\altaffilmark{16,17}, M.~Schultheis\altaffilmark{18}, 
K.~Sellgren\altaffilmark{2}, R.~Terrien\altaffilmark{14,15}, J.~van~Saders\altaffilmark{2}, 
C.~Allende~Prieto\altaffilmark{16,17},
D.~Bizyaev\altaffilmark{19}, A.~Burton\altaffilmark{4}, K.~Cunha\altaffilmark{20,21},
L.~N.~da~Costa\altaffilmark{7,21},
S.~Hasselquist\altaffilmark{4}, F.~Hearty\altaffilmark{4}, 
J.~Holtzman\altaffilmark{22}, A.~E.~Garc\'ia~P\'erez\altaffilmark{4}, 
M.~A.~G.~Maia\altaffilmark{7,21},
R.~W.~O'Connell\altaffilmark{4}, C.~O'Donnell\altaffilmark{4}, M.~Pinsonneault\altaffilmark{2}, 
B.~X.~Santiago\altaffilmark{7,23},
R.~P.~Schiavon\altaffilmark{24}, M.~Shetrone\altaffilmark{25}, V.~Smith\altaffilmark{21,26},
J.~C.~Wilson\altaffilmark{4} 
}
\altaffiltext{1}{NSF Astronomy and Astrophysics Postdoctoral Fellow; gail.zasowski@gmail.com}
\altaffiltext{2}{Department of Astronomy, The Ohio State University, Columbus, OH, 43210, USA}
\altaffiltext{3}{Center for Cosmology and Astro-Particle Physics, The Ohio State University, Columbus, OH, 43210, USA}
\altaffiltext{4}{Department of Astronomy, University of Virginia, Charlottesville, VA, 22904, USA}
\altaffiltext{5}{Department of Physics and Astronomy, Texas Christian University, Fort Worth, TX, 76129, USA}
\altaffiltext{6}{Department of Astronomy, University of Michigan, Ann Arbor, MI, 48109, USA}
\altaffiltext{7}{Laborat\'{o}rio Interinstitucional de e-Astronomia-LIneA, Rio de Janeiro, RJ 20921-400, Brazil}
\altaffiltext{8}{Observat\'{o}rio do Valongo, Universidade Federal do Rio de Janeiro, Rio de Janeiro, RJ 20080-090, Brazil}
\altaffiltext{9}{Osservatorio Astronomico di Padova-INAF, I-35122, Padova, Italy}
\altaffiltext{10}{Yerkes Observatory, The University of Chicago, Williams Bay, WI, 53191, USA}
\altaffiltext{11}{US Naval Observatory, Flagstaff Station, Flagstaff, AZ, 86001, USA}
\altaffiltext{12}{Department of Astrophysical Sciences, Princeton University, Princeton, NJ, 08544, USA}
\altaffiltext{13}{Lowell Observatory, Flagstaff, AZ, 86001, USA}
\altaffiltext{14}{Department of Astronomy \& Astrophysics, The Pennsylvania State University, University Park, PA, 16802, USA}
\altaffiltext{15}{Center for Exoplanets \& Habitable Worlds, The Pennsylvania State University, University Park, PA, 16802, USA}
\altaffiltext{16}{Instituto de Astrof\'{\i}sica de Canarias, Calle V{\'i}a L{\'a}ctea s/n, E-38205 La Laguna, Tenerife, Spain}
\altaffiltext{17}{Departamento de Astrof{\'i}sica, Universidad de La Laguna, E-38206 La Laguna, Tenerife, Spain}
\altaffiltext{18}{Institut Utinam, CNRS UMR 6213, OSU THETA, Universit{\'e} de Franche-Comt{\'e}, 25000 Besan\c{c}on, France}
\altaffiltext{19}{Apache Point Observatory, Sunspot, NM, 88349, USA}
\altaffiltext{20}{Steward Observatory, University of Arizona, Tucson, AZ, 85721, USA}
\altaffiltext{21}{Observat\'{o}rio Nacional, Rio de Janeiro, RJ 20921-400, Brazil}
\altaffiltext{22}{Department of Astronomy, New Mexico State University, Las Cruces, NM, 88003, USA}
\altaffiltext{23}{Instituto de F\'{i}sica, UFRGS, RS 91501-970, Brazil}
\altaffiltext{24}{Astrophysics Research Institute, Liverpool John Moores University, Wirral, CH41 1LD, UK}
\altaffiltext{25}{McDonald Observatory, The University of Texas at Austin, Austin, TX, 78712, USA}
\altaffiltext{26}{National Optical Astronomy Observatories, Tucson, AZ, 85719, USA}

\begin{abstract}

The Apache Point Observatory Galactic Evolution Experiment (APOGEE) is a high-resolution infrared spectroscopic
survey spanning all Galactic environments (i.e., bulge, disk, and halo), with the principal goal of 
constraining dynamical and chemical evolution models of the Milky Way. 
APOGEE takes advantage of the
reduced effects of extinction at infrared wavelengths to observe the inner Galaxy and bulge at an unprecedented level of detail.
The survey's broad spatial and wavelength coverage enables users of APOGEE data to 
address numerous Galactic structure and stellar populations issues.
In this paper we describe the APOGEE targeting scheme and document its various target classes
to provide the necessary background and reference information
to analyze samples of APOGEE data with awareness of the imposed selection criteria and resulting sample properties.
APOGEE's primary sample consists of $\sim$10$^5$ red giant stars, selected to minimize observational biases in
age and metallicity.
We present the methodology and considerations that drive the selection of this sample  
and evaluate the accuracy, efficiency, and caveats of the selection and sampling algorithms.
We also describe additional target classes that contribute to the APOGEE sample,
including numerous ancillary science programs,  
and we outline the targeting data that will be included in the public data releases.

\end{abstract}

\setcounter{footnote}{0}

\section{Introduction} \label{sec:intro}
The Apache Point Observatory Galactic Evolution Experiment (APOGEE) is a near-infrared ($H$-band; 1.51--1.70 $\mu$m), 
high-resolution ($R \sim 22,500$), 
spectroscopic survey targeting primarily red giant (RG) stars
across all
Galactic environments \citep[][and {\it in prep}]{Majewski_2012_apogee}.  
The spectrograph's 
capability to produce 300 simultaneous spectra is facilitated
by many new technologies, such as a system for coupling ``warm'' and cryogenically-embedded fiber optic cables, 
a $30.5 \times 50.8$ cm volume phase holographic grating, 
and a six-element cryogenic camera
focusing light onto three Teledyne H2RG detectors.
See \citet{Wilson_2012_apogee} and Wilson et al., {\it in prep} 
for details of the APOGEE hardware design and construction.
APOGEE is part of the Sloan Digital Sky Survey III \citep[SDSS-III;][]{Eisenstein_11_sdss3overview}, 
observing during bright time on the 2.5-meter Sloan telescope \citep {Gunn_2006_sloantelescope}
at the Apache Point Observatory in Sunspot, NM, USA.
After a commissioning phase spanning May--September 2011, the APOGEE survey officially commenced during
the September 2011 observing run, and observations are expected to continue until the end of SDSS-III in June 2014.

The primary observational goal of the APOGEE survey is to obtain precise and accurate radial velocities (RVs)
and chemical abundances for $\sim$10$^5$ RG stars spanning nearly
all Galactic environments and populations. 
APOGEE targets comprise mostly first-ascent red giant branch (RGB) stars, red clump (RC) stars, and asymptotic giant branch (AGB) stars.
This unprecedented dataset will fulfill several major objectives, in particular:
\begin{itemize}
\item constrain models of the chemical evolution of the Galaxy;
\item constrain kinematical models of the bulge(s), bar(s), disk(s), and halo(s) and discriminate substructures within these components;
\item characterize the chemistry of kinematical substructures in all Galactic components;
\item infer properties of the first generations of Milky Way stars,
through either direct detection of these first stars or measurement of the chemical compositions 
of the most metal-poor stars currently accessible;
\item observe the dust-enshrouded inner Galaxy and bring our understanding of its chemistry and kinematics 
on par with what
is currently available for the solar neighborhood and unobscured halo regions; and
\item provide a statistically significant stellar sample for further investigations into the 
properties of subpopulations or specific Galactic regions.
\end{itemize}

To achieve these objectives, 
the survey's target selection procedures strive to produce a homogeneous, minimally biased sample of RG targets that 
is easily correctable to represent the total underlying giant population in terms of age, chemical abundances, and kinematics.

In this paper, we describe the motivation and technical aspects behind the selection of APOGEE's calibration and science target samples.
\S\ref{sec:gensurvey} contains a summary of the overall survey targeting philosophy, observing strategy, and target documentation.
\S\ref{sec:fieldplan} briefly describes the APOGEE field plan as it pertains to target selection considerations,
and \S\ref{sec:targselection} contains the details of the base photometric catalog 
along with the reddening corrections, color and magnitude limits, and magnitude sampling.  
In \S\ref{sec:redcalib}, we describe the calibration target scheme adopted to aid in overcoming
the challenges imposed by telluric absorption and airglow on ground-based high-resolution IR spectroscopy.  
In \S\ref{sec:eval} we evaluate 
the accuracy and efficiency of our target selection algorithms based on 
data taken during the survey's first year. 
\S\S\ref{sec:clusters}--\ref{sec:addltargs} and Appendix~\ref{sec:anc} contain descriptions of APOGEE's ``special'' targets, such as 
stellar clusters, stellar parameters calibrator targets, and ancillary program targets.  
Finally, in \S\ref{sec:release}, we 
list the targeting and supplementary data that 
will be included along with the first APOGEE data release in SDSS Data Release 10 (DR10).
{\it Readers are strongly encouraged to refer to Appendix~\ref{sec:glossary}, which contains a glossary of SDSS- and APOGEE-specific terminology 
that will be encountered in this paper, other APOGEE technical and scientific papers, and the data releases.}

\section{Survey Targeting and Observation Strategies} \label{sec:gensurvey}

Red giant stars are the most effective tracer population to target for questions of large-scale Galactic structure, dynamics, and chemistry because
they are luminous, ubiquitous, and members of stellar populations with a very wide range of age and metallicity.
Because they are luminous, they can be seen to very great distances, allowing samples of populations far out in the halo and across the disk,
even beyond the bulge.  Because they are ubiquitous, we can observe large numbers of them in all directions, 
allowing for statistically-significant samples even when divided into smaller subsamples by, e.g., Galactic kinematical component or age.
And because RG stars are found in stellar populations of most ages and metallicities, 
we can use them to measure quantitative differences
across these populations and trace their evolution in a Galactic context.

To minimize possible sample biases, the target selection must be based as much as possible on the intrinsic
property distributions of the stars selected.
The observed photometry of stars is determined by the intrinsic stellar properties (such as effective temperature and metallicity)
but is also affected by interstellar extinction, which varies enormously within APOGEE's footprint 
(spanning the Galactic Center to the North Galactic Cap; \S\ref{sec:fieldplan}).
To mitigate these effects, 
the target selection includes reddening corrections.
However, because APOGEE is the first large survey of its type, and because we desire a sample whose selection function is easy to determine,
every effort has been made to minimize the total number of selection criteria,
with particular attention to those that may potentially introduce sample biases with respect to metallicity or age.

\subsection{Overview of APOGEE Observations} \label{sec:observations}

In this section we present a brief description of APOGEE's observation scheme, as an introduction to some
of the most relevant SDSS-III/APOGEE-specific terminology.  
This discussion will be considerably expanded in subsequent sections, 
and Appendix~\ref{sec:glossary} contains a glossary of terms for reference.

The survey uses standard SDSS plugplates, with holes for 300 APOGEE fibers; 
of these, $\sim$70 fibers are reserved for telluric absorption calibrators and airglow emission calibration positions (\S\S\ref{sec:tellurics}--\ref{sec:sky}), 
and the remaining $\sim$230 fibers are placed on science targets.
The patch of sky contained within each plate's field of view is called a ``field'', defined by its central coordinates and angular diameter;
the latter ranges from 1--3$^\circ$, depending on the field's location in the sky (\S\ref{sec:fieldplan}).
The base unit of observation for most purposes is a ``visit'', 
which corresponds to slightly more than one hour of detector integration time.\footnote{Visits 
comprise typically eight individual ``exposures'', which are approximately eight minutes of integration each, 
taken at one of two $\sim$0.5 pixel offset dither positions.
Sub-pixel dithering in the spectral direction is required 
because, at the native detector pixel size, the resolution element is under-sampled in the bluer section of APOGEE spectra.  
These multiple dithered exposures are combined by the data reduction pipeline to produce a single ``visit'' spectrum (Nidever et al.,\ {\it in prep}).}

The number of visits per field varies from one to $\sim$24, for different types of fields (\S\ref{sec:fieldplan}).
Most APOGEE fields are visited at least three times (excluding, e.g., the bulge fields; \S\ref{sec:bulge}) to permit detection of
spectroscopic binaries in the APOGEE sample.  
With typical RV variations of a few ${\rm km}\,{\rm s}^{-1}$ or more, spectroscopic binaries
can complicate the interpretation of APOGEE's kinematical results --- for example, by inflating velocity dispersions.
In addition, given a bright enough companion, the derived stellar parameters may be influenced by the companion's flux,
so the detection of these systems is very useful.
Furthermore, fields with more visits can have samples with fainter magnitude limits (\S\ref{sec:magrange})
that still meet the survey's S/N goal.
Visits are separated by at least one night and may be separated by more than a year, 
depending on the given field's observability and priority relative to others at similar right ascensions.

\begin{figure*} 
\begin{center}
\includegraphics[angle=90,width=0.75\textwidth]{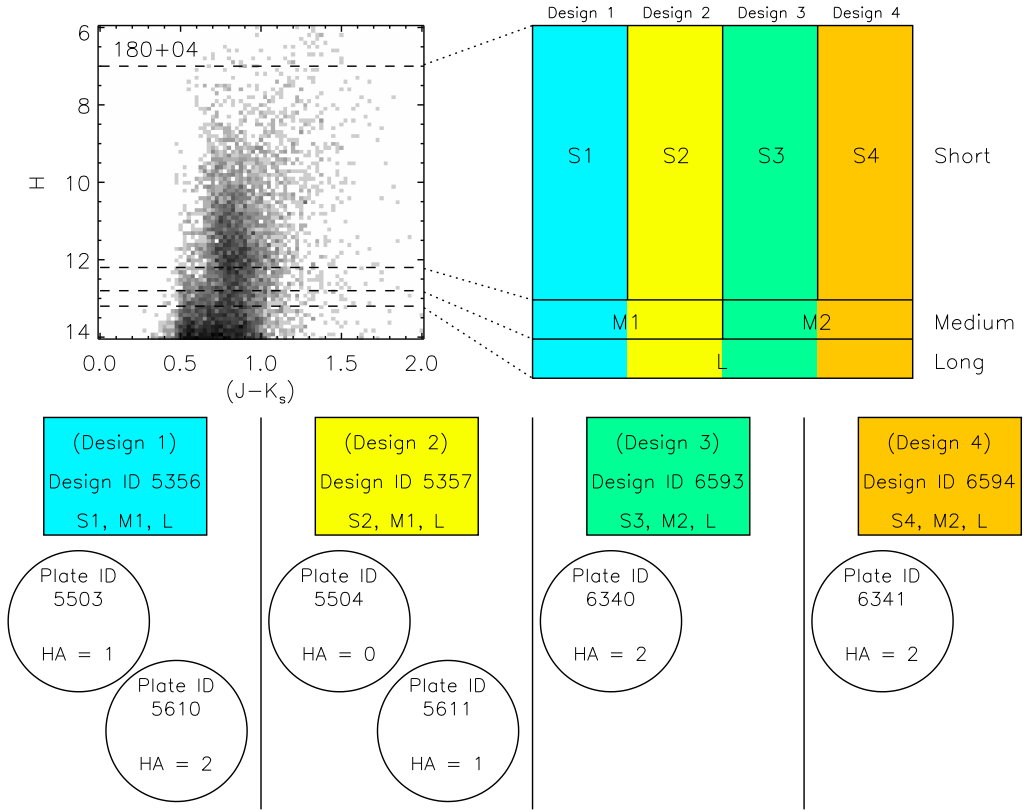}
\end{center}
\caption{
Organization of observed targets in plate designs and on physical plates, using the field {\it 180+04} as an example.  
This field has 12 anticipated visits, 
which are covered by four designs (indicated by blue, yellow, green, and orange).  
Each design has stars from one of four short cohorts (S1, S2, S3, S4), one of two medium cohorts (M1, M2),
and the long cohort (L); that is, stars in the long cohort appear in all four designs, and stars from the medium cohorts appear in two designs.
At least one plate is drilled for each design, and some designs (here, the first two) are drilled on multiple plates.  
Most frequently, this occurs when a field is to be observed at different hour angles (HA), as in this example.
}
\label{fig:cohorts}
\end{figure*}

Different stars may be observed on different visits to a field.  
Stars are grouped into sets called ``cohorts'', based on their $H$-band apparent magnitude,
and each cohort is observed for only as many visits (generally in multiples of three) as needed for all stars in the cohort to achieve
the final desired S/N.  For example, the brightest candidate targets in a given 12-visit field may only need three visits
to reach this goal, whereas stars one magnitude fainter need all 12 visits to reach the same S/N.  Observing
the bright stars for all 12 visits would be an inefficient use of observing time, so a cohort composed of these stars is only
observed three times, and then replaced with another cohort of different bright stars, while a cohort composed of the fainter stars
is observed on all 12 visits to the field.  
Thus, by grouping together cohorts with different magnitude ranges on a series of plates, we
increase the number of total stars observed
without sacrificing stars at the faint end of the APOGEE magnitude range (see additional details on the cohort scheme in \S\ref{sec:magrange}).

A particular combination of cohorts (equivalently, a particular combination of stars) defines a ``design'', with a unique ID number;
a given cohort may appear on a single or on multiple designs.  
See Figure~\ref{fig:cohorts} for an example.
Each physically unique aluminum ``plate'' is drilled with a single design, but a given design may appear on multiple plates ---
for example, if a new plate is drilled for observing the same stars at a different hour angle.
Thus a field (a location on the sky) may have multiple designs (sets of targets), and each design may have multiple plates,
but a plate has only one design, and a design is associated with only one field.
We anticipate $\sim$650 designs to be made over the course of the survey for the approximately 450 distinct fields (\S\ref{sec:fieldplan}).

\subsection{Targeting Flags} \label{sec:targflags}
Reconstruction of the target selection function, however simple it may be, is crucial 
for understanding how well the spectroscopic target sample
represents the underlying population in the field. 
To track the various factors considered in each target's selection and prioritization, APOGEE has defined two 32-bit integers,
{\it apogee\_target1} and {\it apogee\_target2}, whose bits correspond to specific target selection criteria 
(Table~\ref{tab:targflags}).  Every target in a given design is assigned one of each of these integers, 
also called ``targeting flags'' (Appendix~\ref{sec:glossary}),
with one or more bits ``set'' to indicate criteria that were applied to place a target on a design.

These flags indicate selection criteria for a given {\it design}, or particular set of stars (Appendix~\ref{sec:glossary}), 
and thus may differ for the same star on different designs and plates.  
For example, many commissioning plates were observed without a dereddened-color limit (\S\ref{sec:colorrange}), 
so a bit used to indicate that a target was selected because of its dereddened color (e.g., \atone = 3, ``dereddened with RJCE/IRAC'') 
would not be set for those observations; 
however, if later designs drilled for that same field {\it do} have a color limit, and the same stars are re-selected and observed,
that bit {\it would} be set for those later observations of the same stars.

Throughout this paper, we will use the notation \atone = $X$ to indicate that bit ``X'' is set in the \atone flag (and likewise for {\it apogee\_target2}),
even though mathematically, that bit is set by assigning \atone = $2^X$.
Because a target may have multiple ($N$) bits set, its final integer flag value is a 
summation of all set bits: 
\[
\sum\limits_{i=0}^N 2^{\rm bit(i)}.
\]
In keeping with earlier SDSS conventions, if any bit in \atone or \attwo is set, bit 31 for that flag is also set.
For example, a well-studied star that is targeted as a stellar chemical abundance standard (\attwo = 2) 
and also as a member of a calibration cluster (\attwo = 10; see \S\ref{sec:calibclusters}) would have a final
32-bit integer flag of \attwo $= 2^2 + 2^{10} + 2^{31} = -2147482620$
(the negative sign is a result of the fact that these are signed integers).

\begin{deluxetable*}{lc|lc} \tablewidth{0pt}
\tablecaption{Targeting Flags}
\tablehead{ \multicolumn{2}{c}{{\it apogee\_target1}} &  \multicolumn{2}{c}{{\it apogee\_target2}} \\ 
 \colhead{Selection Criterion} & \colhead{bit} & \colhead{Selection Criterion} & \colhead{bit}}
\tablecolumns{4}
\tabletypesize{\footnotesize}
\startdata
--- & 0 & --- & 0 \\
--- & 1 & Flux standard & 1 \\
--- & 2 & Abundance/parameters standard & 2 \\
Dereddened with RJCE/IRAC & 3 & Stellar RV standard & 3 \\
Dereddened with RJCE/WISE & 4 & Sky target & 4 \\
Dereddened with SFD $E(B-V)$ & 5 & --- & 5 \\
No dereddening & 6 & --- & 6 \\
Washington+DDO51 giant & 7 & --- & 7 \\
Washington+DDO51 dwarf & 8 & --- & 8 \\
Probable (open) cluster member & 9 & Telluric calibrator & 9 \\
Extended object & 10 & Calibration cluster member & 10 \\
Short cohort (1--3 visits) & 11 & Galactic Center giant & 11 \\
Medium cohort (3--6 visits) & 12 & Galactic Center supergiant & 12 \\
Long cohort (12--24 visits) & 13 & --- {\it Young Embedded Clusters} & 13 \\
--- & 14 & --- {\it MW Long Bar} & 14 \\
--- & 15 & --- {\it B[e] Stars} & 15 \\
``First Light'' cluster target & 16 & --- {\it Cool} Kepler {\it Dwarfs} & 16 \\
Ancillary program target & 17 & --- {\it Outer Disk Clusters} & 17 \\
--- {\it M31 Globular Clusters} & 18 & --- & 18 \\
--- {\it M Dwarfs} & 19 & --- & 19 \\
--- {\it Stars with High-R Optical Spectra} & 20 & --- & 20 \\
--- {\it Oldest Stars} & 21 & --- & 21 \\
--- Kepler {\it \&} CoRoT {\it Ages} & 22 & --- & 22 \\
--- {\it Eclipsing Binaries} & 23 & --- & 23 \\
--- {\it Pal 1 GC} & 24 & --- & 24 \\
--- {\it Massive Stars} & 25 & --- & 25 \\
Sgr dSph member & 26 & --- & 26 \\
Kepler asteroseismology target & 27 & --- & 27 \\
Kepler planet-host target & 28 & --- & 28 \\
``Faint'' target & 29 & --- & 29 \\
SEGUE sample overlap & 30 & --- & 30 
\enddata
\tablecomments{Bits 13--17 in \attwo also refer to ancillary programs.  
Bits with ``---'' as their criterion have either yet to be defined or were reserved for criteria never applied to released data.}
\label{tab:targflags}
\end{deluxetable*}

\section{APOGEE Field Plan} \label{sec:fieldplan}
\begin{figure*} 
\begin{center}
\includegraphics[angle=90,width=0.9\textwidth]{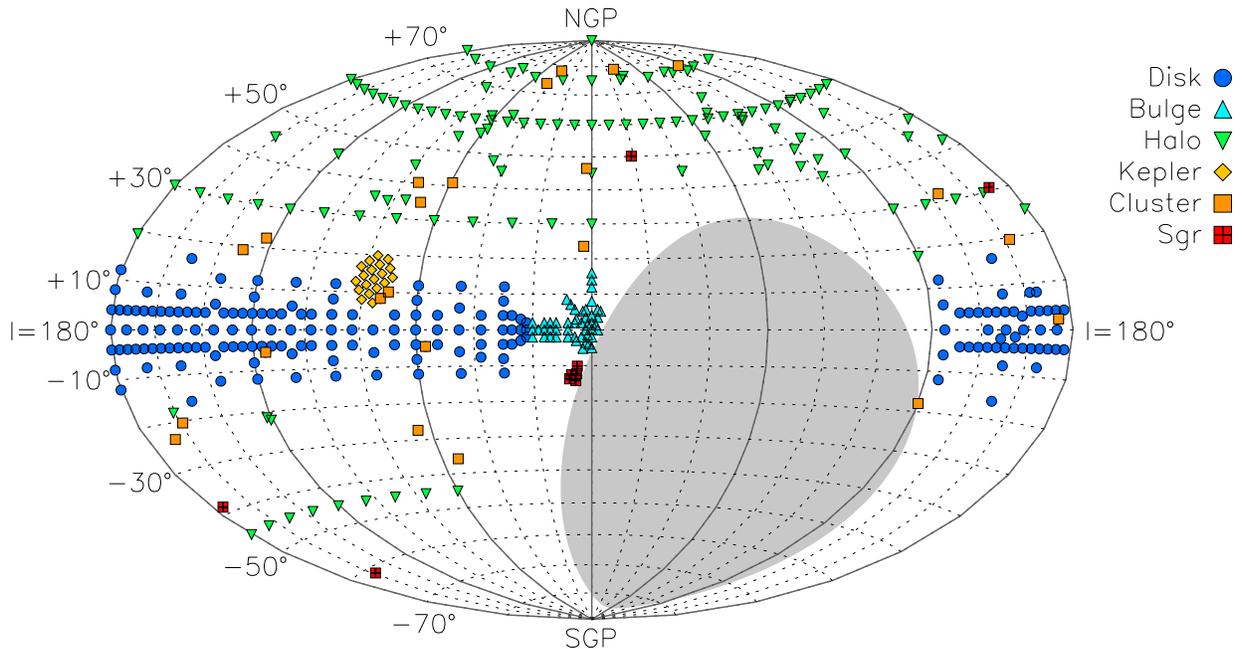}
\end{center}
\caption{Map of the APOGEE field plan.  The map is in Galactic coordinates, with the Galactic Center in the middle,
the anti-center ($l=180^\circ$) on the left and right, and the North/South Galactic Caps at the top and bottom, respectively.
The lines of Galactic latitude are labeled on the left, 
and the solid gray lines indicate Galactic longitudes (from left to right) 
$l = 180^\circ, 120^\circ, 60^\circ, 0^\circ, 300^\circ, 240^\circ$, and $180^\circ$.
The gray shaded area indicates those regions of the Galaxy that are never visible with an airmass $\lesssim$2.3 from APO.
See text for description of the field types.
}
\label{fig:fieldmap}
\end{figure*}

We provide here a summary of the current APOGEE field locations as they pertain to target selection considerations and procedures; 
see Majewski et al., {\it in prep} for a full discussion of the plan's motivation and details.
The APOGEE survey footprint spans as wide a range of the Galaxy as is visible from the Apache Point Observatory (${\rm latitude} = 32.8^\circ$ N), 
and samples all major Galactic components.
Figure~\ref{fig:fieldmap} shows the current complement of chosen field centers (summarized in Table~\ref{tab:fields}). 
``Disk'' fields (\S\ref{sec:disk}) are in dark blue circles, ``bulge'' fields (\S\ref{sec:bulge}) are in light blue point-up triangles, and ``halo'' fields (\S\ref{sec:halo})
are in green point-down triangles.  In addition to these primary classifications, the field plan includes pointings 
covering the footprint of NASA's {\it Kepler} mission (yellow diamonds), well-studied open and globular clusters (orange squares), and 
the Sagittarius dwarf galaxy core and tails (red quartered squares), as described in \S\S\ref{sec:clusters}--\ref{sec:addltargs}.

Most fields are named using the Galactic longitude $l$ and latitude $b$ of their center (i.e., ``{\it lll$\pm$bb}''), 
though we note that these centers are approximate in many cases, and 
the exact coordinates should be obtained from the database if field position accuracy $\lesssim$0.5$^\circ$ is required.
A subset of fields, particularly in the halo, are named for an important object or objects they contain,
such as specific stellar clusters or stellar streams (e.g., the Sagittarius tidal streams; \S\ref{sec:sgr}).  
In these cases, the fields are deliberately {\it not} centered
on the object, because the SDSS plates have a 5 arcmin hole in the center (used to attach the plate to the fiber cartridges) 
that precludes any fiber holes being placed there.
Throughout this paper, we will use {\it italics} when referring to all field names, to remove ambiguity between general 
discussion of targeting in a field
named after a specific object and targeting in the object itself 
(e.g., the APOGEE field pointing {\it M13} versus the globular cluster M13).   

\begin{deluxetable}{lcc} \tablewidth{0 pt} 
\tablecaption{Field Plan Summary}
\tablehead{ \colhead{Type} & \colhead{Definition} & \colhead{Approx. Target Fraction} }
\tabletypesize{\small}
\startdata
Disk & $24^\circ \leq l \leq 240^\circ$, $|b| \leq 16^\circ$ & 50\% \\
Bulge & $357^\circ \leq l \leq 22^\circ$, $|b| \leq 8^\circ$ & 10\% \\
Halo &$|b| > 16^\circ$  & 25\% \\
Special & On calibration/ancillary sources & 15\%
\enddata
\label{tab:fields}
\end{deluxetable}

\subsection{Disk} \label{sec:disk}
The subset of APOGEE fields termed ``disk'' fields form a semi-regular grid spanning $24^\circ \leq l \leq 240^\circ$, with $|b| \leq 16^\circ$.
Each of these fields will be visited 
from 3 to 24 times, meaning that their nominal faint magnitude limits range from $H$ = 12.2 to 13.8 (\S\ref{sec:magrange}).  
For the 3-visit fields, all stars in the selected sample will be observed on all 3 visits, 
while the fields with $>$3 visits employ the cohort scheme described in \S\ref{sec:observations} and 
\S\ref{sec:magrange} to balance the desires for dynamic range, survey depth, and good statistics.  

All stars in the disk grid fields are selected based on their dereddened $(J-K_s)_0$ colors (\S\ref{sec:colorrange}).
Simulations of the survey estimate that 
approximately 50\% of the final survey stellar sample will come from the disk fields (Table~\ref{tab:fields}).
In addition to the normal APOGEE sample and a variety of ancillary targets (Appendix~\ref{sec:anc}), the disk fields contain 
open clusters (\S\ref{sec:openclusters}) falling serendipitously in the survey footprint.

\subsection{Bulge} \label{sec:bulge}
The set of fields considered ``bulge'' fields are those spanning $357^\circ \leq l \leq 22^\circ$ and $|b| \leq 8^\circ$
(plus fields centered on the Sagittarius dwarf galaxy, \S\ref{sec:sgr}).  
Due to the low altitude of these fields at APO,\footnote{For example, the Galactic Center transits the meridian at an altitude of 28$^\circ$.}
and the strong differential atmospheric refraction that results from observing at such high airmasses, 
the bulge fields are restricted to a 1--2$^\circ$ diameter field of view (FOV), compared to the 
full $3^\circ$ diameter for the majority of the survey fields.  
The density of target candidates meeting APOGEE's selection criteria is so high (up to $\sim$7500 deg$^{-2}$), however, 
that even with the restricted FOV, there are ample stars from which to choose in these fields.
Stars in the bulge fields are selected based on their dereddened $(J-K_s)_0$ color, and
approximately 10\% of the final survey sample is projected to come from the bulge fields.

The right ascension (RA) range of the bulge also includes many of the closely-packed inner disk fields.
Because of this RA oversubscription and the small window during which the low-declination bulge can be observed on any given night, 
the majority of the bulge fields are only visited once, instead of the $\geq$3 visits anticipated for all other fields.
The few multi-visit exceptions include high-priority calibration fields, such 
as the Galactic Center, Baade's Window, and those that overlap fields from other surveys \citep[such as BRAVA;][]{Rich_2007_brava}.  
While APOGEE cannot distinguish single-lined spectroscopic binaries in the 1-visit fields, 
it is worth noting that the magnitude limit for these fields ($H \leq 11.0$) 
is still faint enough to include RGB stars in the bulge behind $A(V) \lesssim 25$ mag of extinction.

Special targets in the bulge fields include nearly 200 bulge giants and supergiants, 
already studied with high-resolution optical or IR spectroscopy (\S\ref{sec:gc}). 
These targets are useful for calibrating APOGEE's stellar abundance and parameters pipeline, 
particularly at high metallicity.

\subsection{Halo} \label{sec:halo}
APOGEE's ``halo'' fields are defined as those with $|b| > 16^\circ$, and in practice all have $|b| \geq 18^\circ$.  The stellar population distribution 
in these fields is often substantially different from those of the disk and bulge.
For example, the dwarf-to-giant ratio within APOGEE's nominal color and magnitude range is
much higher in the halo fields, due to the overall lower density of distant giants (see \S\ref{sec:colorrange}).
To improve the selection efficiency of giants, we have acquired additional photometry in the optical Washington $M$ \& $T_2$ and {\it DDO51}
filters \citep[hereafter, ``Washington+DDO51'';][]{Canterna_1976_washfilters,Clark_1979_ddo51filter,Majewski_00_W+Dtech} 
for $\sim$90\% of the halo fields, to assist with identifying and prioritizing giant and dwarf candidates.  
See \S\ref{sec:wddata} for details on the acquisition and reduction of these data.

The paucity of targets in certain halo fields (compared with APOGEE's capability to observe 230 simultaneous science targets)
requires some special accommodations when selecting targets.
One of these is the deliberate targeting of dwarf stars in fields lacking sufficient bright giants (\S\ref{sec:wddata}),
and another is the inclusion of targets with $H$ magnitudes up to 0.8 mags 
fainter than the nominal limits for the fields.  
These ``faint'' targets, which are not expected to attain a final ${\rm S/N} \geq 100$, 
have bit \atone = 29 set and are described more fully in \S\ref{sec:calibclusters}.
In addition, many of the halo fields are placed on open or globular clusters with well-known abundances (\S\ref{sec:calibclusters}), 
and members of these clusters can comprise up to 75\% of all targets in their field.  

Approximately 25\% of the final survey sample is estimated to come from the halo fields.
These survey sample percentages from the different field types do not include the $\sim$15\% 
coming from the ``calibration'' or other special fields, which include the 3-visit bulge fields, 
the long 12--24-visit halo cluster fields (\S\ref{sec:calibclusters}),
and the ``APOGEE--{\it Kepler}'' fields (\S\ref{sec:kepler}).

\section{Photometric Target Selection Criteria and Procedures} \label{sec:targselection}

\subsection{Base Photometric Catalogs and Quality Requirements} \label{sec:quality}
The Two Micron All Sky Survey (2MASS) Point Source Catalog \citep[PSC;][]{Skrutskie_06_2mass} 
forms the base catalog for the targeted sample.
The use of 2MASS confers several advantages:
(i) The need to construct a photometric pre-selection catalog of our own is eliminated.
(ii) The all-sky coverage allows us to draw potential targets 
from a well-tested, homogeneous catalog for every field in the survey.
(iii) Even in the most crowded bulge fields, where, due to confusion, the magnitude limit of the PSC is brighter than in other parts of the 
Galaxy,\footnote{\url{http://www.ipac.caltech.edu/2mass/releases/allsky/doc/sec2\_2.html}}
the PSC is deep enough for APOGEE's nominal magnitude limits.
(iv) The wavelength coverage is well-matched to APOGEE, 
and we can select targets based directly on their $H$-band ($\lambda_{\rm eff} = 1.66\,\mu$m) magnitude.
(v) The astrometric calibration for stars within APOGEE's magnitude range 
is sufficiently accurate (on the order of $\sim$75 mas\textsuperscript{29}) 
for positioning fiber holes in the APOGEE plugplates,
even in closely-packed cluster fields.
Furthermore, the PSC contains merged multi-wavelength photometry 
(the $J$- and $K_s$-bands, with $\lambda_{\rm eff} = 1.24$ and $2.16\,\mu$m, respectively) 
useful for characterizing stars (e.g., with photometric temperatures), as well as detailed data and reduction quality flags for each band.

We combine the 2MASS photometry with mid-IR data to calculate the extinction for each potential stellar target (\S\ref{sec:colorrange}).
Where available, we use data from the {\it Spitzer}-IRAC Galactic Legacy Infrared Mid-Plane Survey Extraordinaire 
\citep[GLIMPSE;][]{Benjamin_03_glimpse,Churchwell_09_glimpses}.
The GLIMPSE-I/II/3D surveys together
span $|b| \lesssim 1^\circ$ for $l \lesssim 65^\circ$ and $l \gtrsim 295^\circ$, with extensions up to $|b| \lesssim 4^\circ$
in the bulge and at select inner-Galaxy longitudes.
Where GLIMPSE is not available, we use data from the all-sky Wide-field Infrared Survey Explorer mission \citep[WISE;][]{Wright_10_WISE};
preference is given to GLIMPSE largely because of {\it Spitzer}-IRAC's higher angular resolution.

To ensure that the colors and magnitudes used in the target selection are accurate 
measurements of the sources' apparent photometric properties, 
we apply the data quality restrictions tabulated in Table~\ref{tab:dataqual} for all potential targets.
These restrictions only apply to the ``normal'' APOGEE target sample; 
ancillary or other special targets (such as calibration cluster members)
are not subject to these requirements.

\begin{deluxetable*}{p{8cm}p{2cm}p{4cm}} \tablewidth{0 pt}
\tablecaption{Adopted Data Quality Criteria For APOGEE Targets}
\tablehead{ \colhead{Parameter} & \colhead{Requirement} & \colhead{Notes} }
\tabletypesize{\small}
\startdata
2MASS total photometry uncertainty for $J$, $H$, and $K_s$ & $\leq$0.1 &   \\
2MASS quality flag for $J$, $H$, and $K_s$ & =`A' or `B' &   \\
Distance to nearest 2MASS source for $J$, $H$, and $K_s$ & $\geq$6 arcsec &   \\
2MASS confusion flag for $J$, $H$, and $K_s$ & =`0' &   \\
2MASS galaxy contamination flag & =`0' &   \\
2MASS read flag & =`1' or `2' &   \\
2MASS extkey ID & {\it null} & For design IDs $\geq$5782  \\
{\it Spitzer} IRAC total photometric uncertainty for $[4.5\mu]$ & $\leq$0.1 & Not strictly enforced on design IDs $\leq$5402\tablenotemark{a} \\
WISE total photometric uncertainty for $[4.5\mu]$ & $\leq$0.1 & No quality limit was imposed on design IDs $\leq$6190. \\
{\it chi} for $M$, $T_2$, and {\it DDO51} data & $<$3 & For design IDs $\geq$5788 \\
$|${\it sharp}$|$ for $M$, $T_2$, and {\it DDO51} data & $<$1 & For design IDs $\geq$5788
\enddata
\tablenotetext{a}{Due to a bookkeeping error, 
sources on some design IDs $\leq$5402 using IRAC data passed the quality check if they either
met the photometric uncertainty requirement in this table {\it or} did not have an IRAC counterpart at $[4.5\mu]$. 
This error appears limited to the commissioning and first 4 designs of {\it 060+00} 
(design IDs 4610, 4820, 4821, 5401, 5402; $\sim$15\% of the stars in those designs), 
the commissioning designs of {\it 006+02} (design IDs 4688, 4689; $\sim$0.5\% of the stars), 
and the single designs of {\it 027+00} and {\it 045+00} (design IDs 5376, 5377; $\sim$15\% of the stars).
Users wishing to recreate accurately the pool of available candidates for these particular designs should be aware of this anomaly.}
\label{tab:dataqual}
\end{deluxetable*}

\subsection{Additional Photometry in Halo Fields} \label{sec:wddata}
As demonstrated in, e.g., \citet{Geisler_1984_lumclass}, \citet{Majewski_00_W+Dtech}, 
\citet{Morrison_2000_mappinghalo}, and \citet{Munoz_05_W+D-UMi}, 
the combination of the Washington and {\it DDO51} filters provides a way to distinguish giant stars from
late-type dwarf stars that have the same broad-band photometric colors.  
The intermediate-band {\it DDO51} filter encompasses the gravity-sensitive Mg triplet and MgH features around 5150 \AA,
and in a $(M-T_2), (M-${\it DDO51}$)$ color-color diagram, 
the low surface gravity giants separate from the high surface gravity dwarfs over a wide range of temperatures.

Our Washington+DDO51 data 
were acquired with the Array Camera on the 1.3-m telescope of the U.S. Naval Observatory, Flagstaff Station.
The Array Camera is a $2 \times 3$ mosaic of $2{\rm k} \times 4{\rm k}$ e2v CCDs, with 0.6$^{\prime\prime}$ pixels
and a FOV of $1.05^\circ \times 1.41^\circ$.  
Each of the APOGEE halo and globular
cluster fields that were observed with the Array Camera was imaged with a pattern of six slightly overlapping pointings.
At each pointing, a single exposure was taken in each of the $M$, $T_2$, and {\it DDO51}
filters, with exposure times of 20, 20, and 200 seconds, respectively, for
non-cluster halo fields, and of 10, 10, and 100 seconds for globular cluster fields.  All
imaging was done under photometric conditions and calibrated against standards
from \citet{Geisler_1990_washstandards,Geisler_1996_washstandards}.

Each image was bias-subtracted, flat field-corrected using sky flats,
and (for the $T_2$ images only) fringing-corrected, using the Image Reduction
and Analysis Facility software \citep[IRAF;][]{Tody_1986_iraf,Tody_1993_irafin90s}.\footnote{IRAF 
is distributed by the National
Optical Astronomy Observatories, which is operated by the Association of
Universities for Research in Astronomy, Inc., under cooperative agreement
with the National Science Foundation.}  
For each pointing, the $M$, $T_2$, and {\it DDO51} images were registered and stacked together.  
Object detection was performed on each stacked image using both SExtractor \citep{Bertin_1996_sextractor}
and DAOPHOT-II \citep{Stetson_1987_daophot}, and the merged detection list was then used
as the source list for the individual images.  DAOPHOT-II was used to
model the point spread function (PSF), which was allowed to vary quadratically
with position in the frame, and to measure both PSF and aperture magnitudes for
each object.  There were positionally dependent systematic differences
between the PSF and aperture magnitudes, which were fit using a quadratic
polynomial as a function of radial distance from the center of the FOV.  While
the residuals around this fit were typically $\sim$0.01 mag, for individual
frames they could be considerably larger and actually comprise the dominant source of photometric error
for those frames.  The raw aperture-corrected PSF magnitudes were then
calibrated against the \citet{Geisler_1990_washstandards,Geisler_1996_washstandards} 
standards using IRAF's PHOTCAL package.  
For most nights, the photometric calibrations yield rms residuals of about 0.02 mag.

\begin{figure} [h!]
\begin{center}
\includegraphics[angle=90,width=0.45\textwidth]{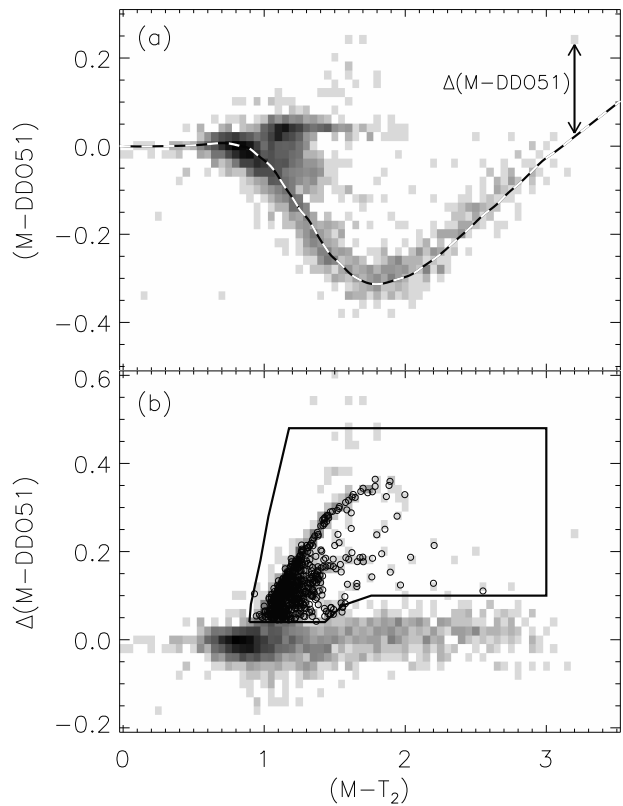} 
\end{center}
\caption{
Demonstration of dwarf/giant separation using Washington+DDO51 photometry.
{\it (a):} $(M-T_2), (M-${\it DDO51}$)$ color-color diagram of stars in the {\it M53} field.
The dashed line indicates the dwarf locus fit for this field, and the vertical arrow on the right
demonstrates how the quantity $\Delta(M-${\it DDO51}$)$ is measured.
{\it (b):} $\Delta(M-${\it DDO51}$)$ as a function of $(M-T_2)$ for the same stars in panel {\it (a)}.
The selection box used to identify giant stars is shown, 
and stars lying within this box that also meet all of APOGEE's data quality criteria are overplotted with open circles.
}
\label{fig:wdlumclass}
\end{figure}

Figure~\ref{fig:wdlumclass} demonstrates the application 
of this Washington+DDO51 photometry to classify giant and dwarf candidates.
First, we defined the shape of the dwarf locus in the $(M-T_2), (M-${\it DDO51}$)$ color-color diagram 
using the full set of stars with good Washington+DDO51 photometry, 
binning the stars in $(M-T_2)$ and iteratively rejecting $(M-${\it DDO51}$)$ outliers in each bin.
Then, separately for each field 
(Figure~\ref{fig:wdlumclass} shows the halo cluster field {\it M53}), 
we ``fit'' this dwarf locus in the
$(M-T_2), (M-${\it DDO51}$)$ color-color diagram (Figure~\ref{fig:wdlumclass}a),
holding the locus shape constant but
allowing small ($\lesssim$0.1 mag) shifts along each axis to account for any residual systematic offsets in the photometry for that field.
Based on the $\Delta(M-${\it DDO51}$)$ distances from this locus as a function of $(M-T_2)$ color,
we then identified the stars likely to be giants using the color-color selection box shown in Figure~\ref{fig:wdlumclass}b.
The minimum and maximum $(M-T_2)$ ``edges'' indicate the colors at which 
the dwarf and giant loci merge for hotter and cooler stars, respectively.

We also used the ``{\it chi}'' and ``{\it sharp}'' values provided by the DAOPHOT-II reduction to gain additional
leverage against non-point source (e.g., cosmic ray or extragalactic) contaminants.  
Only sources  
lying within the ``giant'' color-color selection box, with
$chi < 3$ and $|sharp| < 1$ (where the {\it chi} and {\it sharp} limits are applied to all three bands),
and meeting the additional data quality criteria in Table~\ref{tab:dataqual}
are considered giant target candidates. 
These stars have bit \atone = 7 set.  In the Figure~\ref{fig:wdlumclass}b example, they are overplotted as open circles.
Stars meeting the {\it chi} and {\it sharp} restrictions but classified as ``dwarfs'' (i.e., falling outside the selection box),
have bit \atone = 8 set and
are specifically targeted in some sparse fields lacking sufficient giant candidates to fill all the science fibers
(using the selection and priorities described in \S\ref{sec:calibclusters}).
The accuracy of this classification approach is assessed in \S\ref{sec:eval-wd}.

\subsection{Reddening Estimation and Color Range of Targets} \label{sec:colorrange}

\subsubsection{Application of a $(J-K_s)_0$ Color Limit} \label{sec:colorrangeapp}
To balance the desire for a RG-dominated target sample with the desire for a homogeneous sample
across a wide range of reddening environments, 
the survey's only selection criterion (apart from magnitude) is a single color limit 
applied to the {\it dereddened} $(J-K_s)_0$ color.
To derive the extinction corrections, we use the Rayleigh Jeans Color Excess
method \citep[RJCE;][]{Majewski_11_RJCE}, which calculates reddening values 
on a star-by-star basis using a combination of near- and mid-IR photometry.
As described in \S\ref{sec:quality}, the near-IR data come from the 2MASS PSC, 
and we use mid-IR data from the {\it Spitzer}-IRAC GLIMPSE-I/II/3D and WISE surveys (\atone = 3 and 4, respectively). 
Specifically, here we use the $H$ and 4.5 $\mu$m data:
\begin{eqnarray} \label{equ:rjce}
A(K_s) & = & 0.918 \times (H - [4.5\mu] - (H-[4.5\mu])_0) \\  
E(J-K_s) & = & 1.5 \times A(K_s), \nonumber
\end{eqnarray}
where $A(K_s)$/$E(J-K_s)$ is adopted from \citet{Indeb_05_extlaw}.  
We adopt $(H-[4.5\mu])_0 = 0.08$ for all IRAC data \citep{Girardi_02_isochrones} and, 
after a comparison of IRAC and preliminary WISE 4.5 $\mu$m photometry in the midplane, $(H-[4.5\mu])_0 = 0.05$ for all WISE data.

\begin{figure*} 
\begin{center}
\includegraphics[angle=90,width=0.9\textwidth]{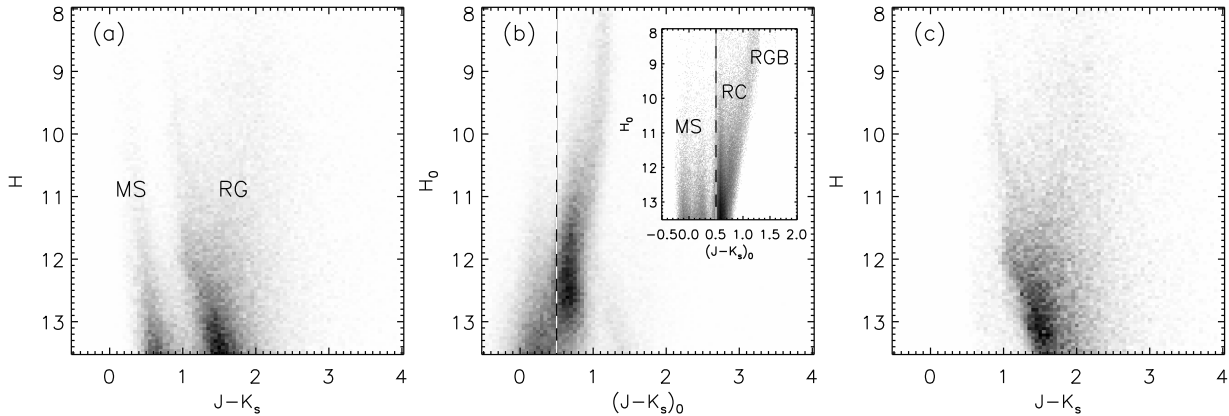} 
\end{center}
\caption{Demonstration of the effects of the adopted dereddened color limit.
{\it (a):} Uncorrected 2MASS CMD of all of the stars in the {\it 060+00} field meeting the survey photometric quality criteria.
``MS'' and ``RG'' indicate the regions of the CMD dominated by main sequence and red giant stars, respectively.
{\it (b):} RJCE-corrected CMD of the same stars.  
The inset shows an extinction-free TRILEGAL stellar populations simulation of this same field \citep{Girardi_05_trilegal}.
The dashed lines indicate $(J-K_s)_0=0.5$, the color limit adopted for APOGEE's giant star sample (\S\ref{sec:colorrangejust}).
{\it (c):} Uncorrected CMD of the stars meeting the dereddened color requirement.  
Note that the broad diagonal swath of main sequence stars has been preferentially removed.
}
\label{fig:drcmds}
\end{figure*}

Figure~\ref{fig:drcmds} demonstrates the application of this dereddened color limit, using the {\it 060+00} field as an example.
Figure~\ref{fig:drcmds}a is the observed 2MASS color-magnitude diagram (CMD) of the stars 
meeting the data quality criteria given in Table~\ref{tab:dataqual}.  
Note the broad locus of MS stars extending from $(J-K_s) \sim 0.3$ at $H \sim 10.5$ to $(J-K_s) \sim 0.7$ at $H \sim 13.5$,
and the much wider swath of RG stars spanning $1 \lesssim (J-K_s) \lesssim 3$ for nearly the full range of $H$ shown.
Though it may be relatively easy to distinguish the two loci visually here, the properties and shape of the gap between them (when it exists)
depends very strongly on the field's distribution of reddening, 
and a very complex algorithm would be required to select the giant stars to the red side of the gap in uncorrected CMDs
across the wide variety of stellar populations and reddening environments contained within APOGEE's fields.  

A much simpler and homogeneous approach is to 
apply an {\it intrinsic} color limit across the entire survey.
In Figure~\ref{fig:drcmds}b, we show the reddening-corrected CMD for this field;
in the inset is a simulated CMD of the same field center and size but with zero extinction, 
drawn from the TRILEGAL Galactic stellar populations model \citep[][see \S\ref{sec:colorrangejust}]{Girardi_05_trilegal}.
The vertical dashed lines in both CMDs denote the $(J-K_s)_0 \geq 0.5$ color limit adopted for APOGEE's ``normal'' targets. 
The choice of this particular limit is described in \S\ref{sec:colorrangejust} below, but
Figure~\ref{fig:drcmds}c shows the result of its application.  The data are identical to Figure~\ref{fig:drcmds}a,
except that only stars meeting the color limit are shown.  
Nearly all of the MS stars, and almost none of the RG stars, have been removed from the sample,
which demonstrates the effectiveness of this technique at preferentially targeting giant stars regardless of the reddening properties of a given field.

Evaluation of the first year of survey data revealed a systematic {\it over}-correction of many of the halo targets,
which was partly traced to a metallicity dependence --- specifically, low-metallicity stars (${\rm [Fe/H]} \lesssim -1.1$) have redder
$(H-[4.5\mu])_0$ colors than more metal-rich ones, leading to an overcorrection for metal-poor stars, 
which reside preferentially in the halo fields (see further details in \S\ref{sec:eval-dr}).  
Rather than adopting a field-specific intrinsic color (in effect, assuming a mean [Fe/H] as a function of $l,b$),
we chose to use the integrated Galactic reddening maps of \citet[][hereafter ``SFD'']{Schlegel_98_dustmap} as an upper limit on the reddening
towards stars in the halo fields.  That is, we adopt
\begin{equation}
A(K_s) = 0.302 \times E(B-V)_{\rm SFD},
\end{equation}
for each star 
for which the $E(J-K_s)$ value calculated from the star's photometry using Equation~\ref{equ:rjce}
is greater than 1.2$\times$ the SFD-derived value.
The conversion between $E(B-V)_{\rm SFD}$ and $E(J-K_s)$ is taken from \citet{Schlafly_2011_calibSFD}, and
the factor of 1.2 is used to provide a margin of tolerance, based on the 
typical photometric uncertainty, when comparing the two reddening values.

This ``hybrid'' dereddening method (so called because stars in the same design can be selected with different dereddening techniques)
is applied only to 3-visit fields in the halo, with $|b| \geq 16^\circ$ and design ID 6919 or later.  
Halo fields with more than 3 visits (i.e., those with multiple designs) 
are excluded because at least some of the designs had already been drilled during the first year 
of survey operations, and we elected to preserve the homogeneity of the target selection across all designs for a given field.
Disk and bulge fields are excluded for a number of reasons.
First, the SFD map values are not applicable in the midplane and in regions of high extinction or with steep extinction gradients
\citep[e.g., SFD;][]{Arce_1999_testSFD98,Chen_1999_extinctionmodel}.
Second, we have verified that the vast majority of the observed stars in these fields are in 
fact correctly dereddened with the RJCE method alone (\S\ref{sec:eval-dr}).
Finally, most of these fields are part of a deliberate grid pattern, 
with corresponding fields across key symmetry axes (such as the midplane) already observed during the first year;
therefore, we elected not to adopt this change to the targeting algorithm that would reduce the 
grid's selection homogeneity while not actually improving the target selection efficiency.

In the end, then,
a simple dereddened color selection of $(J-K_s)_0 \geq 0.5$ is applied for most normal targets in the survey.
For the well-populated bulge and disk fields, we require a non-null and positive
extinction estimate (i.e., $A[K_s] \geq 0$),\footnote{Because 
the near-IR 2MASS catalog is the base catalog for the survey, this requirement translates to a requirement of a 
mid-IR detection.  APOGEE's magnitude ranges are within the completeness limit for both the IRAC and WISE surveys, 
so we expect nearly all non-detections
in the mid-IR data to be due to data issues in those surveys (such as proximity to bright, very red stars) that 
do not impose an intrinsic-property bias
on the final sample.}\textsuperscript{,}\footnote{For exceptions, see Note $a$ in Table~\ref{tab:dataqual}.}
but for the sparse halo fields, the target density is low enough that to fill all 230 science fibers on a plate, 
we often include targets without an extinction estimate, simply requiring an observed $(J-K_s) \geq 0.5$.
The exceptions are the 3-visit halo fields selected with the hybrid dereddening scheme described above; 
in these designs, the SFD map value is used in place of any missing RJCE-WISE values.

The homogeneity and simplicity of the color selection adopted here should allow for a straightforward
reconstruction of the selection function and evaluation of any biases in the final target sample,
which --- in large part because of this approach --- we expect to be very minor.

\subsubsection{Justification of the Adopted $(J-K_s)_0$ Color Limit} \label{sec:colorrangejust}
Our choice of a color cut at $(J-K_s)_0 \geq 0.5$ was motivated by two main
considerations: (i) to include stars cool enough for a
reliable derivation of stellar parameters and abundances via the 
APOGEE Stellar Parameters and Chemical Abundances Pipeline (ASPCAP; Garcia Perez et al., {\it in prep}),
and (ii) to keep the fraction of nearby dwarf star ``contaminants'' in the sample as low as possible.

Both observational data and theoretical isochrones demonstrate that
dwarfs and giants of the same \teff span nearly identical ranges of NIR color for $(J-K_s)_0 \lesssim 0.8$.
Solar metallicity M dwarfs of subtype $\sim$M5 or earlier have a maximum color of $(J-K_s)_0 \sim 0.85$
\citep[][]{Koornneef_1983_NIRcolors2,Bessell_1988_jhklm,Girardi_02_isochrones,Sarajedini_2009_msturndown}. 
Other dwarf stellar objects --- 
e.g., heavily-reddened M dwarfs, M dwarfs of subtypes later than M5 
\citep[e.g., Table 2 of][]{Scandariato_2012_lowmassNIRcolorsOrion}, or brown dwarfs --- 
may reach colors redder than this, but these populations are extremely rare at the magnitudes relevant for APOGEE. 
A simple color limit of $(J-K_s)_0 \geq 0.85$ would therefore
eliminate the vast majority of potential dwarf contaminants from the survey sample. 
However, this criterion would also eliminate the RC giants, which for near-solar
metallicities concentrate at $(J-K_s)_0 \sim 0.5-0.7$, along with the more metal-poor RG stars. 
RC stars are highly desirable targets for APOGEE
due to their high density among the total MW giant population
and nearly constant absolute magnitude (making them effective ``standard candles'').
A color limit of $(J-K_s)_0 \geq 0.85$ would also
restrict the sample to the coolest giants ($T_{\rm eff} \lesssim 4300$ K, for solar
metallicity), leading to a strong bias towards high metallicities
and subjecting the survey to the systematically greater uncertainties that plague abundance analyses of very cool giants.

Therefore, a color limit bluer than $(J-K_s)_0=0.85$ was sought, 
and we adopted $(J-K_s)_0 \geq 0.5$ as the primary criterion for selecting ``normal'' APOGEE targets
after exploring the following quantitative considerations about the expected dwarf star fraction in APOGEE's magnitude range.

To estimate the dwarf fraction in the 2MASS catalog as a function of $(l,b)$, 
we utilized the TRILEGAL model \citep{Girardi_05_trilegal,Girardi_2012_trilegal}, a population synthesis model of the Galaxy that
simulates complete samples of stars along pencil beam lines of sight, 
including all of the stellar properties needed for these tests (such as multi-band photometry and surface gravities);
the model includes an approximation of 2MASS photometric errors 
and is able to reproduce the 2MASS star counts to the $\sim$20\% level in low-reddening regions.
We performed extensive TRILEGAL simulations of the original APOGEE field plan,
assuming: (i) a thin disk with a total mass surface density of 55.4 $M_\sun pc^{-2}$, a scalelength $h_R = 2.9$ kpc, 
and an age-dependent scaleheight $h_Z(t_{\rm Gyr})=94.7(1+t/5.5)^{1.66}$ pc;
(ii) a thick disk with a local mass volume density of 10$^{-3} M_\sun pc^{-3}$, $h_R = 2.4$ kpc, and $h_Z=0.8$ kpc;
and (iii) a halo, modeled as an oblate $R^{1/4}$ spheroid, with a local mass volume density of 10$^{-4} M_\sun pc^{-3}$,
an oblateness of 0.58 in the $Z$ direction, and a semi-major axis of 2.7 kpc.  
These parameters are the default values for the improved version of TRILEGAL described in \citet{Girardi_2012_trilegal}.
Mass densities are computed assuming a \citet{Chabrier_2001_imf} initial mass function, 
and the age-metallicity distribution of the halo and disk components is described in \citet{Girardi_05_trilegal}.
(TRILEGAL also includes a bulge, but it does not impact the fields for which results are shown below.)

\begin{figure}[!h]
\includegraphics[width=0.35\textwidth,angle=90]{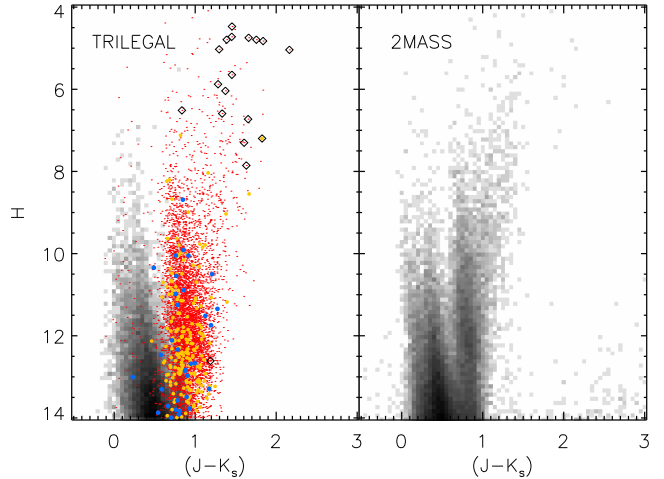} 
\caption{
CMDs for a 7 deg$^2$ field centered at $(l,b) = (180^\circ,3^\circ)$ from
a TRILEGAL simulation (left) and 2MASS (right).
In the simulation, the grayscale density plot represents dwarfs, and colored symbols are giants:
red dots for stars in the thin disk, yellow circles for thick disk, and blue circles for halo.
Thermally-pulsing AGB stars are marked with open diamonds.
}
 \label{fig_cmd2mass}
\end{figure}

The TRILEGAL simulations clearly indicate that, for stars redder than $(J-K_s)_0=0.5$,
the dwarf fraction, defined as the fraction of stars with $\log g>3.5$,
increases both with increasing apparent magnitude and
towards the Galactic poles. 
These trends in the dwarf fraction result from the decreasing numbers of distant giants being sampled at high Galactic latitudes, 
as well as from the large difference in absolute magnitude between cool giants and cool dwarfs ---
cool giants are intrinsically bright enough that even ones located in the distant MW halo have apparent magnitudes
brighter than many nearby cool dwarfs.
In the low-latitude example of Figure~\ref{fig_cmd2mass}, the dwarf fraction for stars with $(J-K_s)_0>0.5$
and $8<H<11$ is $\sim$3\%, increasing to $\sim$9\% at $11<H<12$.  
We consider these fractions acceptable for a survey
like APOGEE, and these estimates are overall quite reassuring,
especially when considering that the dwarf fraction further
decreases towards the low-latitude, inner-Galaxy regions that contain the majority of APOGEE's pointings, 
and that most fields at higher $|b|$ have supplementary photometry to reduce the 
effects of the increased dwarf contamination there (\S\ref{sec:wddata}).

\begin{figure}
\begin{center}
\includegraphics[width=0.4\textwidth,angle=90]{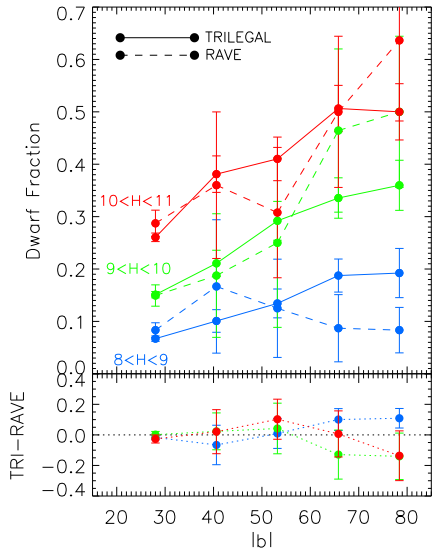}
\end{center}
\caption{
Comparison of the dwarf fractions as a function of $|b|$, in the TRILEGAL model (solid lines)
and RAVE data (dashed lines), for three different ranges of $H$ magnitude, as indicated.
Both the simulated and observed data have been limited to stars with $0.5 \leq (J-K_s) \leq 0.8$,
lying within a strip spanning $40^\circ \leq l \leq 80^\circ$ and $-20^\circ \leq b \leq -90^\circ$.
The bottom panel contains the difference between the TRILEGAL and RAVE dwarf fractions,
also as a function of $|b|$.
}
 \label{fig_strip}
\end{figure}

As a check of these calculations prior to the development of ASPCAP, 
we compared the TRILEGAL dwarf/giant ratio predictions 
to the observed $\log g$ distributions from the RAdial Velocity Experiment \citep[RAVE;][]{Steinmetz_2006_raveDR1}.
RAVE has observed large sections of the Southern sky at $|b|>20^\circ$, 
collecting spectroscopy of $\sim$$200-400$ stars in
each $\sim$28~deg$^2$ survey field position.
Inspection of the 2MASS photometry for RAVE
targets suggests that, within the limits of $0.5 \leq (J-K_s)_0 \leq 0.8$ and $H<10$,
this sample is representative of the underlying stellar distribution. 
We binned the RAVE sample in
small boxes in the $([J-K_s]_0$, $H$) CMD and determined the dwarf
fraction in each box using the $\log g$ values returned from the RAVE
pipeline \citep{Zwitter_2008_raveDR2}; then we repeated the procedure for the TRILEGAL simulations.  
Figure~\ref{fig_strip} shows the dwarf fraction in the $0.5 \leq (J-K_s)_0 \leq 0.8$ interval 
(which contains most of the dwarf contamination expected in APOGEE)
for a series of RAVE and TRILEGAL pointings, averaged over a
$\Delta l = 40^\circ$ strip across the sky and extending from the Southern
Galactic Pole up to $b\sim-20^\circ$. The data and
model predictions for the dwarf fraction agree very well within the error bars.

This comparison, together with the TRILEGAL simulations at lower
latitudes, give us confidence that, with the adopted $(J-K_s)_0 \geq 0.5$ color
limit, the {\it mean} dwarf fraction (considering the full distribution of magnitude) 
will be smaller than 40\% in even the deepest APOGEE plates. 
A forthcoming paper (Girardi et al., {\it in prep.}) will examine the trends between $(l,b)$ and the dwarf/giant ratio in more
detail using additional $\log g$ data, including those from ASPCAP and {\it Kepler}.

An additional prediction of interest for the APOGEE sample that can be extracted from the TRILEGAL
simulations is the fraction of the stellar sample anticipated to be
thermally pulsing asymptotic giant branch (TP-AGB) stars.  These stars are among the most intrinsically luminous
in the Galaxy, but APOGEE will observe many in the heavily extinguished bulge and inner disk that are too faint to be accessible 
by optical spectrographs.  The simulation shown in Figure~\ref{fig_cmd2mass} predicts that 
TP-AGB stars will comprise on the order of 1\% of the stellar sample
with $(J-K_s)_0 \geq 0.5$ and $H \leq 12$.  Though only a single line of sight is shown here, 
this fraction is estimated to be roughly constant (i.e., at the few percent level) throughout the survey footprint.
For these estimations, TRILEGAL uses the TP-AGB evolutionary tracks of \citet{Marigo_2007_tp-agbmodels},
which have lifetimes calibrated on observations of AGB stars in Magellanic Cloud star clusters.
See \S\ref{sec:gc} for a description of the set of known AGB stars deliberately targeted in the bulge.

\subsection{Cohorts and Magnitude Ranges} \label{sec:magrange}

APOGEE's goal of exploring all Galactic populations requires sampling magnitude ranges broad enough
to probe stars at a wide range of distances along the line of sight, as well as stars at a wide
range of intrinsic luminosities. 
To achieve the desired chemical abundance precision ($\leq$0.1 dex), the nominal signal-to-noise (S/N) goal for all APOGEE science
targets is $\geq$100 per {\it pixel} (i.e., ${\rm S/N} \sim 150$ per resolution element for 2-pixel sampling of the line spread profile).
Commissioning data demonstrated that this goal can be achieved for targets with
$H \leq 11.0$ in approximately one hour, which is the length of time for a single visit to a field,
and the ${\rm S/N} = 100$ magnitude limit for the more common three-visit fields is $H \leq 12.2$.

To reach even fainter magnitudes, which probe greater distances as well as 
intrinsically fainter giants,
stars must be visited additional times to build up signal.
Thus many fields are visited more than three times, with some having up to 24 visits 
planned.\footnote{Many of these ``long'' fields, with 6--24 visits, were originally designed to 
accommodate the required observing cadence of the MARVELS survey \citep[\S\ref{sec:marvels};][]{Ge_2008_marvels}.}
As described in \S\ref{sec:observations}, the division of stars into ``cohorts'' permits stars of very different magnitudes
to be observed for different numbers of visits to increase efficiency.

In this scheme, stars observed only three times together are referred to
as a ``short'' cohort, stars observed six times form a ``medium'' cohort, 
and stars observed 12 to 24 times form the ``long'' cohort of their field.
See Table~\ref{tab:maglimits} for the magnitude limits of each cohort type. 
The small number of exceptions to this scheme arise from certain fields with a high fraction of bright calibrator targets or complex observing needs.
One example is the Galactic Center field {\it GALCEN}, which is split into three one-visit short cohorts and one three-visit medium cohort, 
to maximize the number of valuable calibrator stars observed (\S\ref{sec:gc}).  These exceptions are also noted in Table~\ref{tab:maglimits}.

\begin{deluxetable*}{p{1cm}p{1cm}p{13cm}} \tablewidth{0 pt}
\tablecaption{Field and Cohort Magnitude Limits}
\tablehead{ \colhead{$N_{\rm visits}$} & \colhead{$H$-band Limit\tablenotemark{a}} & \colhead{Notes} }
\tabletypesize{\small}
\startdata
1 & 11.0 & most {\it Kepler} fields, most bulge fields, Sgr core fields  \\
 & & \\
3 & 12.2 & ``short'' cohorts in long fields, short disk/halo fields, {\it Kepler} cluster field {\it N6791}, ``medium'' cohorts in the bulge calibration fields {\it GALCEN, BAADEWIN,} and {\it BRAVAFREE} \\
 & & \\
6 & 12.8 & ``medium'' cohorts in long fields, {\it N5634SGR2}, {\it 221+84}, {\it Kepler} cluster field {\it N6819}, MARVELS shared fields {\it N4147} and {\it N5466} \\
 & & \\
12 & 13.3 & ``long'' cohorts in most long disk/halo fields \\
 & & \\
24 & 13.8 & ``long'' cohorts in the longest disk/halo fields: {\it 030+00, 060+00, 090+00}, {\it PAL1}, and {\it M15}
\enddata
\tablenotetext{a}{Apparent magnitude limit for normal APOGEE science targets; ancillary and other special targets are not necessarily restricted by these limits.}
\label{tab:maglimits}
\end{deluxetable*}

One notable consequence of this many-visit scheme is that APOGEE stars are not necessarily 
uniquely identified by a single ``plate--MJD--fiber ID'' combination, as many previous SDSS targets have been.
Such a combination instead identifies a single {\it visit spectrum} that is combined with spectra of the same star
from other visits to produce the final stellar spectrum. 

The saturation limit of the detectors, combined with an unexpected superpersistence problem on 
regions of two of the three detector arrays (Nidever et al.\,{\it in prep}; Wilson et al.\,{\it in prep}), 
have led us to impose a bright limit of $H \geq 7.0$ for science targets, 
extending up to $H \geq 5.0$ only for some of the valuable telluric calibrators (\S\ref{sec:tellurics}).

However, a large number of very valuable calibrator and ancillary targets are brighter than these limits,
sometimes significantly so.
A fiber link between the APOGEE instrument and the NMSU 1-meter telescope at APO \citep{Holtzman_2010_nmsu1m}
was completed during Fall 2012, providing an opportunity to observe very bright targets (e.g., Arcturus),
other targets that are useful for calibration but do not fall within existing APOGEE fields, and
targets needing repeated visits for time series and variability studies (e.g., pulsating AGB stars). 
These 1-meter observations can be made during dark time when
APOGEE is not scheduled for the Sloan 2.5-meter telescope.

\subsection{Magnitude Sampling} \label{sec:magsample}
The final magnitude distribution of an APOGEE design differs from a purely random sampling of the apparent
magnitude distribution in the field due to two factors: (i) the number of fibers allotted to each cohort and 
(ii) the algorithm used to select the final targets from the full set of stars meeting 
the photometric quality, color, and magnitude criteria described in \S\S\ref{sec:quality}--\ref{sec:magrange}.

First, the number of fibers assigned to each cohort is determined as a function of the field's $l$ and $b$,
not by the number of stars available for each cohort.  
For example, in the low-latitude inner disk fields ($l \leq 90^\circ$, $|b| \leq 4^\circ$), 95 fibers are allotted to the long cohorts, 
whereas in the higher-latitude disk fields ($|b| = 8^\circ$), only 30 fibers are reserved for long cohort targets.  
This apportionment was governed by 
expectations of  
whether apparently fainter stars were more likely to be intrinsically fainter or simply farther away 
(the latter being more desirable from a Galactic structure point of view), 
the dwarf/giant ratio as a function of $H$ magnitude, and the thin/thick disk ratio as a function of $H$, all for particular ranges of $l$ and $b$.
Another factor in the fiber allotment is the desire for a large number of targets, 
given that each long cohort star may take the place of up to eight short
cohort stars. 

The cohort fiber allotments are shown in Table~\ref{tab:fiberallocate}, but note that these are approximations --- 
due to other plate design considerations, such as fiber collisions,\footnote{Each of the APOGEE 
fibers are enclosed in a protective stainless steel ferrule with a 71.5 arcsec diameter;
these ferrules prevent stars that are less than 71.5 arcsec apart from being targeted in the same design.} 
the actual number of stars in each cohort on a given design may
differ (generally, by $\pm\lesssim$5).  
Furthermore, these allotments are only valid for the 12- and 24-visit disk fields; other fields with multiple cohorts,
such as the long halo fields, have allotments governed by the distribution of special targets (such as cluster members) within them.

\begin{deluxetable}{ccc} \tablewidth{0 pt}
\tablecaption{Disk Field Cohort Fiber Allocations}
\tablehead{ \colhead{$l$} & \colhead{$b$} & \colhead{Short, Medium, Long Fibers} }
\tabletypesize{\small}
\startdata
$l \leq 90$ & $|b| \leq 4$ & 90, 45, 95\\
$l > 90$ & $|b| \leq 4$ & 90, 90, 50\\
All & $|b| \geq 8$ & 130, 70, 30 \\
\enddata
\tablecomments{Fields with medium and long cohorts that are not in this disk grid have fiber allocations 
determined by the distribution of high-priority targets within the field.}
\label{tab:fiberallocate}
\end{deluxetable}

Second, 
rather than drawing the cohort targets randomly from all candidate targets, 
we attempt to sample stars spaced more evenly in apparent magnitude.  
This is accomplished by first sorting the stars by apparent $H$ mag and then 
dividing them into three bins within each cohort's magnitude range, 
such that each bin contains 1/3 of the available stars for that cohort.  
(That is, each cohort is drawn from three magnitude bins, so that designs with only a short cohort
will have three bins, but designs with a short, a medium, and a long cohort will have nine bins total.)  

Then, for the short cohorts, each {\it bin} is sampled randomly for 1/3 of the desired number of stars 
for that particular cohort.  
For the medium and long cohorts,\footnote{The difference in sampling algorithms between the short and medium/long
cohorts results from the overlap between the commissioning and ``survey'' plate design timelines.  The former, containing only short cohorts,
used the semi-random selection, which was also applied to the short cohorts of the first survey fields containing long cohorts.
The APOGEE team chose to continue this scheme for all of the survey's fields, deciding that discrepancies between
the short and medium/long cohorts of all designs in all fields will be easier to account for 
than discrepancies between short and medium/long cohorts of specific designs in some fields.}
the stars are selected by 
drawing every $N^{\rm th}$ star in the trio of magnitude-sorted bins, where $N$ is defined by the number of stars available for the cohort
and the number of fibers assigned to that particular cohort.  For example, if 1000 stars were available for a cohort, and 100 fibers assigned, the
final cohort would include every 10$^{\rm th}$ star --- i.e., the \{1$^{\rm st}$, 11$^{\rm th}$, 21$^{\rm st}$, \ldots\} stars, when sorted by magnitude.  
These final targets are then prioritized in random order before actually being assigned for drilling on the plate,
to avoid preferring brighter stars within the cohort in the case of fiber collisions.

The type of cohort to which a star is assigned is reflected in its final targeting bitmask (Table~\ref{tab:targflags}), 
where {\it apogee\_target1} = 11 indicates a short cohort, 12 a medium cohort, and 13 a long cohort.

The goal of this sampling algorithm, as compared to a random draw, 
is a brightness distribution sampling less dependent on the variety of intrinsic magnitude distributions across the wide
variety of Galactic environments.  
Because fainter stars are more common, 
this scheme imposes a slight bias towards the brighter end of the magnitude distribution,
by requiring that at least one star be drawn from the brightest 1/3 of the stars within a given cohort.

\begin{figure} [h!]
\begin{center}
\includegraphics[width=0.45\textwidth]{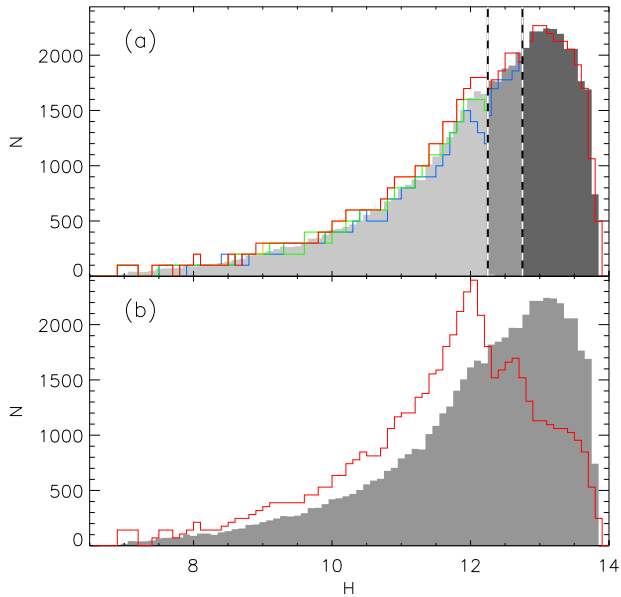} 
\end{center}
\caption{Demonstration of the effects of APOGEE's magnitude sampling, for the disk field {\it 060+00}. 
This is a 24-visit field, 
so the limiting magnitude is $H = 13.8$.  All ancillary, cluster, and other ``special'' targets have been removed.
{\it (a):} The shaded gray histogram shows the apparent magnitude distribution of stars meeting all 
quality, color, and magnitude selection criteria.  Light/medium/dark gray indicate the stars that could have been assigned 
to the short/medium/long cohort(s), respectively, and the dashed lines indicate the magnitude limits of the cohorts.  
The overplotted colored lines show the (vertically stretched) magnitude distribution of stars in cohorts that have
been targeted in this field: four short (blue, orange, green, red), two medium (blue, red), and one long (red).
{\it (b):} The shaded gray histogram is again the apparent magnitude distribution of all available stars in this field, 
and the red line shows the total (vertically stretched) distribution of the stars that have been targeted ---
the summation of the cohorts in panel {\it (a)}. 
Note that while each {\it cohort's} sampling closely approximates its underlying magnitude
distribution, the overall sampling is strongly biased toward brighter stars, especially those near the faint limit of the short cohort.  
See text (\S\ref{sec:magsample}) for additional details.
}
\label{fig:magdist}
\end{figure}

However, upon comparison of the available and selected magnitude distributions,
we find that these sampling procedures produce a final targeted magnitude distribution 
that very closely resembles a random selection {\it within each cohort}.
The top panel of Figure~\ref{fig:magdist} demonstrates this for four designs in the {\it 060+00} field.  All four of these designs share
the same long cohort, two share one medium cohort and two share the other, and all four have unique short cohorts.  
The shaded gray histogram is the apparent $H$ mag distribution of all target candidates, 
and the colored lines show the (vertically stretched) magnitude distributions of the individual short, medium, and long cohorts.
Note that within each magnitude span (i.e., $7.0 < H < 12.2$ for short, $12.2 < H < 12.8$ for medium, and $12.8 < H < 13.8$ for long),
the shape of the cohorts' $H$ distributions very closely resemble that of the underlying population.

Obviously, however, a strong bias would be imposed by not accounting for the effects of combining cohorts of different lengths, 
as demonstrated by the bottom panel of Figure~\ref{fig:magdist}.  
Here, the final targeted magnitude distribution (red line) does {\it not} mimic that of the underlying population,
due to the mismatch between fiber allotment and field magnitude distribution, and further enhanced by the summation of 
multiple short and medium cohorts.
Thus a proper correction for the sampling over the full magnitude range of a field
should account for the three bin divisions within each cohort and the $N^{\rm th}$ sampling in the medium and long cohorts, 
as well as for the distribution of the $\sim$230 science fibers among the short, medium, and long cohorts.

One approach to dealing with this non-random sampling (even including ancillary or other special targets) 
is to compare directly the final targeted sample's color and magnitude distribution with that of the pool from
which it was drawn via the algorithm described above.  (For the vast majority of APOGEE's ``normal'' target sample,
$(J-K_s)_0$ and $H$ are the only parameters used in the selection.)  Then, each spectroscopic target is assigned 
a ``weight'' based on how well its color and magnitude reflect those of the underlying population.  For example, a bias towards
brighter stars (as described above) will manifest itself in a higher fraction of brighter spectroscopic targets than is observed
in the candidate target pool; down-weighting those over-represented targets will prevent the final derived property distribution
(e.g., [Fe/H] or RV) from being skewed towards those targets.  This is in essence the procedure explored by
\citet{Schlesinger_2012_segueGKdwarfs} in their analysis of the [Fe/H] distribution of the SEGUE cool dwarf sample,
which has a much more complex selection function than the APOGEE one described here.

\subsection{Overlap with MARVELS Target Sample} \label{sec:marvels}
For a number of designs observed during Year 1 of APOGEE (through Spring 2012),
a small additional color-magnitude bias in the final target sample was imposed as a result of sharing telescope time
with the Multi-Object APO Radial Velocity Exoplanet Large-area Survey 
\citep[MARVELS;][]{Ge_2008_marvels,Eisenstein_11_sdss3overview}, when plates were observed with fibers
running to the MARVELS and APOGEE spectrographs simultaneously.
The MARVELS targets were selected using proper motions and optical/NIR photometry 
\citep[Paegert et al.,\,{\it in prep}; \S2 of][]{Lee_2011_marvels1b}
but typically inhabit the $0.3 \lesssim (J-K_s) \lesssim 0.9$
and $5 \lesssim H \lesssim 12$ ranges of 2MASS color-magnitude space.
On co-observed plates, the MARVELS targets were prioritized after the APOGEE telluric calibrators (\S\ref{sec:tellurics})
but before the APOGEE science targets; thus APOGEE science target candidates falling within the MARVELS
color-magnitude selection box had a chance, particularly in the sparser halo fields, of being selected
as a MARVELS target and made unavailable to APOGEE.  
Table~\ref{tab:marvels} lists those fields and designs whose plates were drilled for both APOGEE and MARVELS fibers,
using bold text to indicate those that are intended for observation (i.e., not supplanted by APOGEE-only designs).

\begin{deluxetable*}{ll|ll} \tablewidth{0 pt}
\tablecaption{Shared APOGEE+MARVELS Designs}
\tablehead{ \colhead{Field Name} & \colhead{Design ID(s)} & \colhead{Field Name} & \colhead{Design ID(s)} }
\tabletypesize{\footnotesize}
\startdata
 {\it 030+00} &  3959,  {\bf 4810,  4811}	& {\it 180+00} & 2031,  2034 \\
 {\it 030+04} &  {\bf 3961,  4814,  4815} & {\it 180+04}   & 2030 \\
 {\it 030+08} &  3963,  {\bf 4818,  4819} & {\it 180-08}   & {\bf 4860,  4861} \\
 {\it 030-04} &  3960,  {\bf 4532,  4812,  4813}	& {\it 210+00}   & 3235 \\
 {\it 030-08} &  3962,  {\bf 4530,  4816,  4817} & {\it 210+04}   & 3236 \\
 {\it 060+00} &  {\bf 4610,  4820,  4821}  & {\it 210+08}   & 3253, {\bf 5415,  5416,  5417,  5418} \\
 {\it 060+04} &  3965,  {\bf 4538,  4824,  4825} & {\it 210-04}   & 3234 \\
 {\it 060+08} &  3967,  {\bf 4828,  4829}   & {\it 210-08}   & 3229 \\
 {\it 060-04} &  3964,  {\bf 4537,  4822,  4823} & {\it HD46375}  & {\bf 5411,  5412,  5413, 5414 }\\
 {\it 060-08} &  3966,  {\bf 4826,  4827 }  & {\it M107}  & 3233,  3250, {\bf 5784,  5785,  5786,  5787} \\
 {\it 090+00} &  4619,  {\bf 4830,  4831} & {\it M13}  & 3232,  {\bf 3251,  3252},  4696,  4697, {\bf 5782,  5783} \\
 {\it 090+04} &  {\bf 4531,  4834,  4835} & {\it M15}  & 4534,  {\bf 4862,  4863} \\
 {\it 090+08} &  {\bf 4611,  4612},  4613,  4614, {\bf 4838,  4839} & {\it M3}  & 2027, {\bf 3246,  3247,  5482,  5483 }\\
 {\it 090-04} &  {\bf 4533,  4832,  4833 }  & {\it M53}  & 1942,  3231, {\bf 3245,  5484,  5485 }\\
 {\it 090-08} &  4615,  4616,  4617,  4618,  {\bf 4836,  4837} & {\it M5PAL5}   & 3239, {\bf 3248},  3249 \\
 {\it 120+00} &  4628,  {\bf 4840,  4841}    & {\it M92}  & {\bf 3258 }\\
 {\it 120+04} &  4624,  4625,  4626,  4627,  {\bf 4844,  4845} & {\it N2420}  & 1940,  3230,  3242, {\bf 5421,  5422,  5423,  5424 }\\
 {\it 120+08} &  4620,  4621,  4622,  4623,  {\bf 4848,  4849} & {\it N4147}  & 1941,  3244, {\bf 5444,  5445 }\\
 {\it 120-04} &  {\bf 4842,  4843}   & {\it N5466}  & 2028, {\bf 3256,  5486,  5487} \\
 {\it 120-08} &  {\bf 4846,  4847} & {\it N5634SGR2}& 3238, {\bf 3257,  4529 }\\
 {\it 150+00} &  2029,  {\bf 4850,  4851} & {\it N6229}  & 3240 \\
 {\it 150+04} &  {\bf 4852,  4853 }  & {\it NGP}  & 3968 \\
 {\it 150+08} &  {\bf 4856,  4857 } & {\it PAL1}  & {\bf 4864,  4865} \\
 {\it 150-04} &  2030,  2033 & {\it SGR1}  & 2031,  3243, {\bf 5419,  5420} \\
 {\it 150-08} &  {\bf 4854,  4855 }	& {\it VOD1}  & 3237,  3254,  4528, {\bf 5446,  5447,  5448,  5449} \\
 {\it 165-04} &  {\bf 4858,  4859 }	& {\it VOD2}  & 3969,  4535, {\bf 5474,  5475,  5476,  5477}
\enddata
\label{tab:marvels}
\end{deluxetable*}

\section{Atmospheric Contamination Calibration Targets} \label{sec:redcalib}
Despite the many advantages conferred by observing in the near-IR,
two significant spectral contaminants strongly affect this wavelength regime: 
terrestrial atmospheric absorption (``telluric'') lines and airglow emission lines. 
Of the 300 APOGEE fibers observed on each plate, $\sim$35 are devoted to stellar targets used to trace telluric absorption,
and $\sim$35 to ``empty sky'' positions to sample atmospheric airglow.  
(Note that some of the plates designed for commissioning 
observations had different numbers of telluric and sky targets --- 25, 45, or 150 of each --- used to test the number of calibrator fibers needed.) 
Corrections for these contaminants are calculated for all stellar targets in a field by spatially interpolating the 
contamination observed in the calibrator sources across the field (Nidever et al., {\it in prep}).

\subsection{Telluric Absorption Calibrator Targets} \label{sec:tellurics}
\begin{figure} [h!]
\begin{center}
\includegraphics[angle=90,width=0.5\textwidth]{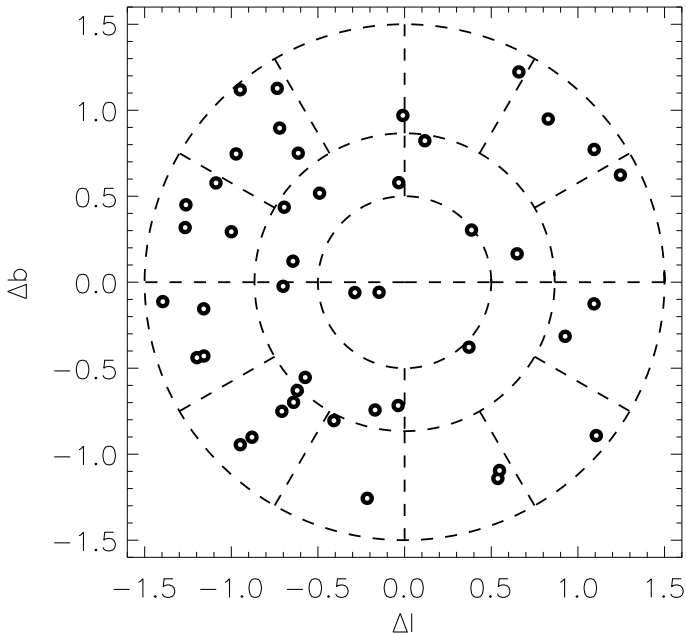} 
\end{center}
\caption{The use of field ``zones'' in the selection of telluric standards (\S\ref{sec:tellurics}) and sky fibers (\S\ref{sec:sky}),
using the field {\it 060+00} as an example.  
The dotted lines demarcate the 18 zones used in the selection of the 44 telluric standard candidates 
(empty black circles; total number includes a 25\% overfill pool to reach 35 final targets).  
We select the bluest star in each zone and then add the $N$ bluest stars remaining in the field, regardless of position, 
where $N$ is the needed remainder (in this case, 26 stars).
For the sky fibers, eight ``empty'' positions randomly selected from each zone form the pool 
from which the final 35 sky fiber positions are drawn.
}
\label{fig:tellsky}
\end{figure}

In the wavelength span of APOGEE, the primary telluric absorption contamination comes from H$_2$O, CO$_2$, and CH$_4$ lines, 
with typical equivalent widths of $\sim$160~m\AA.  
The ideal calibrator targets for dividing out such contamination would be perfect
featureless blackbodies; 
to approximate this situation, 
we select $\sim$35 of the bluest (thus hopefully hottest) stars in each field to serve as telluric calibrators.
Given the $\lesssim$7 deg$^2$ plugplate FOV and $\sim$1-hour integration duration of the individual visits, 
care must be taken to account for both the temporal and spatial variations
in the telluric absorption across the field. 
The temporal variations are incorporated
by observing the telluric calibrators simultaneously with the science targets,
and the spatial variations are monitored by selecting telluric calibrators as follows:

The FOV of each field is divided into a number of segmented, equal-area zones, with the number
of zones being approximately half the number of desired calibrators (see Figure~\ref{fig:tellsky}).  
In each zone, the star with the bluest color (uncorrected for reddening) is 
selected, which ensures that intrinsically red sources with possibly
overestimated reddening values (\S\ref{sec:eval-dr}) are not included in the sample.  
The second half of the calibrator sample, plus a $\sim$25\% overfill, 
is composed of the bluest stars remaining in the candidate pool, regardless of position in the field.  
(Telluric calibrator candidates are subject to the same photometric quality requirements as the science target candidates.)
This dual-step process ensures that almost all of the telluric calibrators will come from
the bluest stars available, but also that they will not be entirely concentrated in one region of the plate 
(due to, say, an open cluster or a random overdensity of blue stars in the field).
No red color limit is imposed on this calibrator sample. 
The telluric calibrator targets chosen in this way have bit \attwo = 9 set, 
and they are prioritized above all science and ``sky'' targets.

We note that observations of these hot stellar targets are producing a unique subsample of high-resolution, near-IR spectra of O, B, and A stars, 
with potential for very interesting science beyond APOGEE's primary goals (e.g., Appendices \ref{sec:massivestars} and \ref{sec:bestars}). 

\subsection{Sky Calibrator Targets} \label{sec:sky}
In addition to telluric absorption, {\it emissive} spectral contamination is contributed by 
IR airglow lines (primarily due to OH), scattered light
from the Moon and light pollution, unresolved starlight, and zodiacal dust.
We dedicate $\sim$35 fibers per plate to ``empty'' positions that are chosen as representative 
of the sky background in the science target fibers for the given field.

The pool of candidate ``sky'' calibrator positions for each field is 
created by generating a test grid of positions spanning the entire FOV of the field
(with grid spacing $\sim$1/2 the fiber collision limit), 
and then comparing each position to the entire 2MASS PSC to calculate the distance of the nearest stellar neighbor.  
Only positions meeting the same ``nearest neighbor'' criterion applied to the science target candidates 
($\geq$6 arcsec; \S\ref{sec:quality}) are considered as candidates.
The positions are not prioritized or sorted by nearest-neighbor distance.

After the pool of candidate positions is generated, the final target list is selected in a method somewhat 
similar to that of the telluric calibrators described above (\S\ref{sec:tellurics}).
The FOV is divided into the same number of zones as used for the telluric standards (Figure~\ref{fig:tellsky}), 
and candidates are drawn from each zone to ensure relatively even coverage of the background
spatial variations.  In this case, however, up to eight candidates are randomly selected from each zone, to ensure
sufficient available targets, and the final list of submitted target
positions is randomly prioritized after the telluric and science targets. 
All sky position ``targets'' have bit \attwo = 4 set.

We emphasize that these sky spectra  
are used to produce maps of atmospheric airglow with high spectral, temporal, and angular resolution with every single observation.
Though APOGEE is using these data simply for calibration purposes, 
they could be used to extract a wealth of information on the physical conditions, chemical composition, 
and variability of Earth's atmosphere itself.  

\section{Evaluation of Target Selection Accuracy and Efficiency With Year 1 Data} \label{sec:eval}
In this section, we assess the performance of the two primary target selection criteria (other than magnitude): 
Washington+DDO51 dwarf/giant classification (\S\ref{sec:eval-wd}) and the dereddened $(J-K_s)_0$ color limit (\S\ref{sec:eval-dr}).
Our goal was to determine to what extent these procedures are producing the desired target sample and ascertain what changes, if any,
needed to be made to improve accuracy and efficiency in Years 2--3 of APOGEE. 
These evaluations of the target selection algorithms are based on
the spectral reductions and derived stellar parameters that comprised a nearly-final version of the DR10 dataset.
We have removed stars with total ${\rm S/N} < 50$  
and stars with reported $T_{\rm eff} \leq 3600$ K or $T_{\rm eff} \geq 5800$ K, where the stellar parameter calculations are
strongly affected by this pipeline version's stellar parameter grid limits. 

\subsection{Washington+DDO51 Giant/Dwarf Separation} \label{sec:eval-wd}
As described in \S\ref{sec:wddata}, Washington+DDO51 photometry can provide additional leverage for the classification of dwarf and giant stars, 
which is particularly useful for efficient targeting in the dwarf-dominated halo fields.  Here, we assess the reliability of this classification algorithm.

\begin{figure*} 
\begin{center}
\includegraphics[angle=90,width=0.9\textwidth]{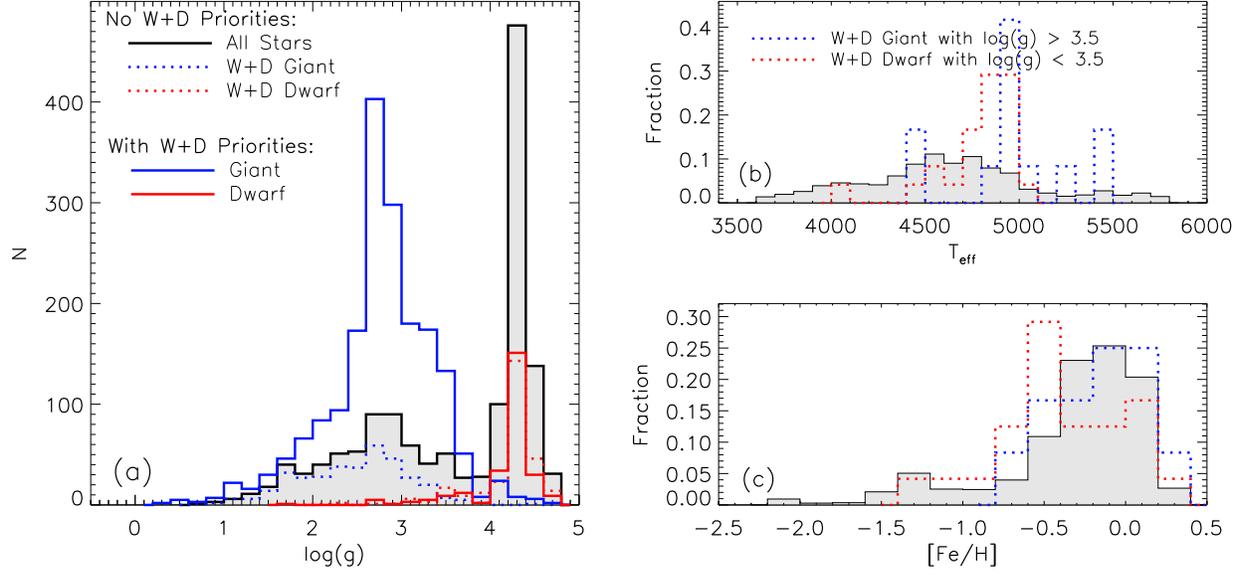} 
\end{center}
\caption{Distributions of ASPCAP log~$g$ values for stars in halo fields ($b \geq 18^\circ$)
observed during APOGEE's first year of operations. 
{\it (a):} Log~$g$ values for stars targeted without reference to a Washington+DDO51-derived 
luminosity class classification (shaded histogram),
the subsets of stars classified as Washington+DDO51 giants (blue dotted line) and dwarfs (red dotted line) but not targeted as such,
and stars intentionally targeted as Washington+DDO51 giants (blue solid line) and dwarfs (red solid line).
{\it (b):} The distribution of \teff values for stars incorrectly classified as dwarfs or giants, adopting $\log g = 3.5$ as
a discriminator.  The blue line represents stars classified as Washington+DDO51 giants but with dwarf-like ASPCAP gravities (i.e., $\log g > 3.5$),
and the red line indicates stars classified as Washington+DDO51 dwarfs but with giant-like ASPCAP gravities (i.e., $\log g < 3.5$).
The shaded histogram contains the same stars as the shaded one in {\it (a)},
and each of the three lines shows the fractional distribution.
{\it (c):} Similar to {\it (b)} but for [Fe/H].
}
\label{fig:wdlogg}
\end{figure*}

In Figure~\ref{fig:wdlogg}a, we show the distribution of ASPCAP log~$g$ values for stars classified as giants and as
dwarfs using Washington+DDO51 photometry, observed in 
32 halo fields ($b \geq 18^\circ$) during the first year of APOGEE operations.
(After application of the S/N and ASPCAP parameter limits described above, these stars comprise $\sim$65\% of the total sample observed
in these fields as of October 2012.) 
The black line, shaded histogram represents stars that were not {\it targeted} as either Washington+DDO51-classified giants or dwarfs.
This category comprises stars that do not have Washington+DDO51 photometry 
meeting the quality requirements described in \S\ref{sec:halo} ---
fainter stars, stars in fields completely lacking Washington+DDO51 data, and
stars falling beyond the edges of the Array Camera CCD chips (\S\ref{sec:wddata}) ---  
as well as stars {\it with} Washington+DDO51 classifications 
but observed on designs that did not incorporate any selection or prioritization using those classifications.  
The blue and red dotted-line distributions explicitly indicate these latter Washington+DDO51 giant and dwarf subsamples.
The shaded distribution demonstrates that the halo is indeed heavily populated by dwarfs within APOGEE's magnitude range, 
and the distinct peaks of the Washington+DDO51-classified giant/dwarf $\log~g$ distributions 
indicate the method's ability to separate the populations relatively cleanly.

The solid blue and red lines in Figure~\ref{fig:wdlogg}a represent the $\log~g$ distributions of stars
that were deliberately targeted as Washington+DDO51 giants and dwarfs, respectively, 
using the prioritization described in \S\ref{sec:calibclusters} (basically, all giants before dwarfs).
We include these distributions to show the significantly higher fraction of giants among the photometrically selected sample,
compared to that among the non-photometrically selected field sample. 

For the combined data of these fields, and using a value of log~$g = 3.5$ as determined by ASPCAP to discriminate giants and dwarfs,
we find that $\sim$4.2\% of the Washington+DDO51 ``giants'' are actually dwarfs and 
$\sim$27\% of the Washington+DDO51 ``dwarfs'' are actually giants.  
In Figures~\ref{fig:wdlogg}b and c, we show the \teff and [Fe/H] distributions of these ``misclassified'' stars (blue and red lines),
along with the distribution of these properties for stars either without a Washington+DDO51 luminosity class or targeted
with no reference to that class (shaded histogram, same stars in the shaded histogram in Figure~\ref{fig:wdlogg}a).

The majority of the ``misclassified'' stars have $4750 \lesssim T_{\rm eff} \lesssim 5000$~K,
corresponding to the range of $(M-T_2)$ colors where the dwarf and giant loci increasingly overlap (Figure~\ref{fig:wdlumclass});
inspection of the $(M-T_2), (M-\mathit{DDO51})$ color-color diagram reveals that nearly all of the rest
lie very close to the ``dwarf locus'' for their field (\S\ref{sec:wddata}).
Figure~\ref{fig:wdlogg}c contains the [Fe/H] distribution for these same stars.
Comparison to the underlying mean population (shaded histogram) demonstrates that the small fraction of luminosity
classification errors does not significantly bias the final targeted sample in metallicity.

\subsection{Dereddening and $(J-K_s)_0$ Color Criteria} \label{sec:eval-dr}
The intrinsic color limit imposed on the survey ($[J-K_s]_0 \geq 0.5$) 
has been made to reduce bias against metal-poor giants (\S\ref{sec:colorrange});
the color of the giant branch at the level of the horizontal branch for a solar metallicity isochrone is $(J-K_s)_0 \sim 0.73$,
while stars at the same evolutionary stage with ${\rm [Fe/H]} \sim -1.3$ have $(J-K_s)_0 \sim 0.53$ \citep[][]{Girardi_02_isochrones}.
Here, we evaluate the accuracy of the dereddened color selection --- i.e., whether the spectroscopic
$T_{\rm eff}$ distribution matches what is predicted by the dereddened $(J-K_s)_0$ distribution.

\begin{figure*} 
\begin{center}
\includegraphics[angle=90,width=0.8\textwidth]{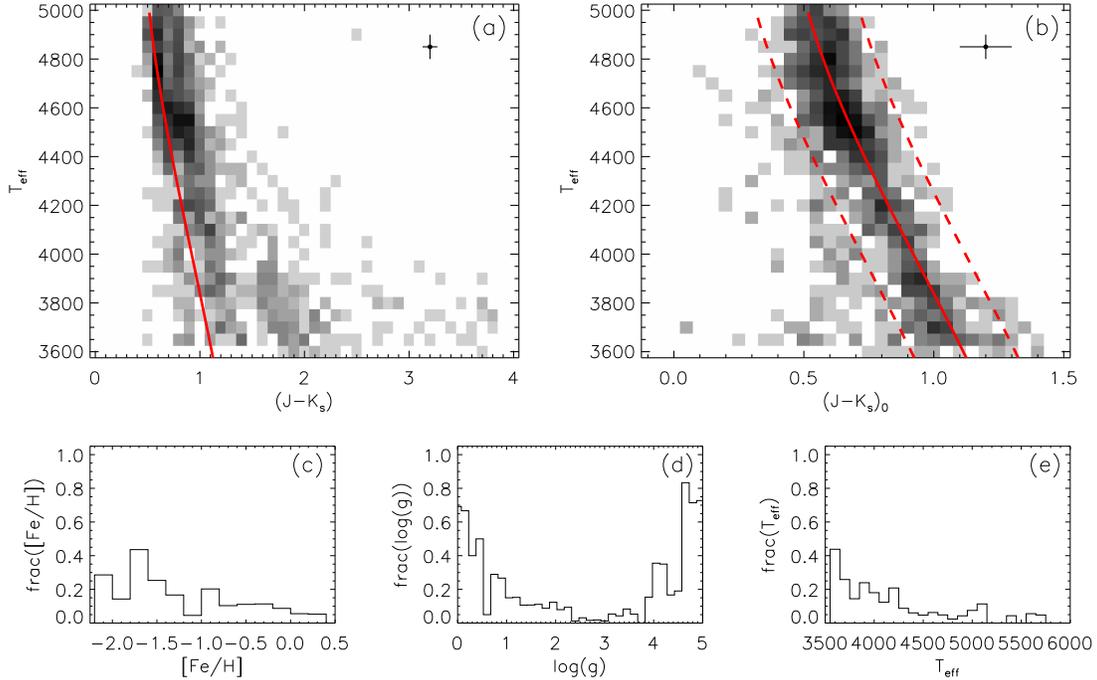} 
\end{center}
\caption{
Comparison between NIR colors and ASPCAP $T_{\rm eff}$ values for stars observed in 13 bulge, disk, and halo fields 
during APOGEE commissioning and Year 1.
{\it (a):} Uncorrected $(J-K_s)$ colors of the stars.  Because this sample comprises fields probing both
the heavily-reddened bulge (down to $l,b = 0^\circ,0^\circ$) and the low-reddening halo (up to $|b| = 56^\circ$), the range of
observed colors is broad.  The solid red line is the locus of color-temperature spanned by giant star isochrones
\citep[$-1.5 \leq {\rm [Fe/H]} \leq +0.2$;][]{Girardi_02_isochrones}.  The typical color 
and \teff uncertainties are shown by the errorbars in the upper right corner.
{\it (b):} Almost identical to {\it (a)}, except for RJCE-corrected $(J-K_s)_0$ colors instead of observed $(J-K_s)$.  
The representative color uncertainty now includes uncertainties from dereddening.
The dashed red lines to either side of the solid line represent the $\sim$1$\sigma$ range for which we consider stars to be in good agreement with the
theoretical color-temperature relationship (see text).
{\it (c):} Distribution of [Fe/H] values for stars {\it not} in the range of good agreement indicated in {\it (b)}, scaled bin-by-bin by the distribution of
[Fe/H] for all stars in this sample.
{\it (d):} Similar to {\it (c)} but for stellar $\log{g}$ values.
{\it (e):} Similar to {\it (c)} but for stellar \teff values.
}
\label{fig:dered}
\end{figure*}

In the top two panels of Figure~\ref{fig:dered}, we directly compare the uncorrected $(J-K_s)$ and 
RJCE-corrected $(J-K_s)_0$ colors to the ASPCAP-derived 
spectroscopic $T_{\rm eff}$ values for stars in 13 fields --- spanning bulge, disk, and halo environments ---
that were observed during APOGEE commissioning and Year 1
({\it GALCEN, 004+00, 000+06, 010+00, 010+02, 014+02, 060+04, 090+04, 090--08, M13, M71, SGRC3}, and {\it VOD3}).
The left-hand panel (Figure~\ref{fig:dered}a) shows the range of uncorrected $(J-K_s)$ colors observed in these fields,
where the wide variety of reddening environments produces a wide range of reddened colors, $0.5 \lesssim (J-K_s) \lesssim 4$.
The right-hand panel (Figure~\ref{fig:dered}b) shows the much narrower RJCE-corrected color range (note the reduced abscissa scale).
In both plots, the solid red line indicates the mean color-temperature relation for giant star isochrones spanning
a range of metallicities \citep[$-1.5 \leq {\rm [Fe/H]} \leq +0.2$; from][]{Girardi_02_isochrones}.  In Figure~\ref{fig:dered}b, the dotted red lines
to either side of this relationship represent a zone of ``reasonable agreement'', after considering the intrinsic range of 
$(J-K_s)_0$ for the set of isochrones at a given $T_{\rm eff}$ combined with the typical uncertainties of the stellar $(J-K_s)_0$ values.
These typical uncertainties are shown in the upper right-hand corner of the upper panels.

These panels demonstrate that, by and large, the RJCE dereddening method performs very well at recovering the 
intrinsic $(J-K_s)_0$ color associated with the spectroscopic $T_{\rm eff}$ for each star.
In Figures~\ref{fig:dered}c, \ref{fig:dered}d, and \ref{fig:dered}e, 
we show distributions of stellar parameters ([Fe/H], log~$g$, and $T_{\rm eff}$, respectively)
for stars lying outside the zone indicated by the dashed red lines in Figure~\ref{fig:dered}b, 
which comprise $\lesssim$15\% of the stars shown.  
The ordinate axis is the number of ``mis-corrected'' stars in each parameter bin normalized by the total number of stars in that bin. 
Clearly, stars in the following ranges of parameter space are most likely to be reddening-corrected {\it away} from the 
theoretical color-temperature relation: low metallicity (${\rm [Fe/H]} \lesssim -1.1$), 
very high or very low surface gravity ($\log{g} \lesssim 0.5$; $\log{g} \gtrsim 4.5$),
and low temperature ($T_{\rm eff} \lesssim 4000$ K).
Some fraction of these apparent outliers is likely due to inaccuracies in the ASPCAP results, which may be correlated;
for example, a giant star assigned an erroneously cool \teff may also be assigned an erroneously low $\log{g}$.
Beyond these issues (which will be improved in future versions of ASPCAP), 
the observed behavior may be due to one or more of the following:

(i) The trend for mis-corrected stars to be
more metal-poor suggests that stars with ${\rm [Fe/H]} \lesssim -1.1$ do not meet RJCE's specific assumptions of color 
homogeneity.  We examined theoretical stellar colors --- specifically, the $(H-[4.5\mu])_0$ color used by the RJCE method (\S\ref{sec:colorrange}) ---
and found that, starting around ${\rm [Fe/H]} \sim -1.3$, the predicted stellar color does indeed increase with decreasing metallicity,
which qualitatively would produce an offset in the direction observed in Figure~\ref{fig:dered}b.
This effect was not observed or discussed by \citet{Majewski_11_RJCE} in their establishment of the RJCE method
because in the inner Galactic midplane fields that were the focus of that work's calibration and analysis, 
the mean stellar metallicity is high enough that the assumption of a common $(H-[4.5\mu])_0$ color is valid.

However, we note that the amount of overcorrection predicted by theoretical colors is insufficient to explain
the full range of offset observed.  For example, a star with ${\rm [Fe/H]} \sim -1.5$ is expected to have 
$(H-[4.5\mu])_0 \sim 0.09$, a difference in color of 0.04 from that assumed for a halo star with WISE data, 
corresponding to a $\Delta (J-K_s)_0 \sim 0.06$ (Equation~\ref{equ:rjce}).  Some stars have $\Delta (J-K_s)_0$ 
of several tenths of a magnitude, so another (perhaps additional) factor is affecting the observed distribution.
Nevertheless, because this overcorrection may remove desirable targets from our sample, 
particularly in the lower-metallicity halo fields,
we have adopted the SFD reddening maps in certain fields as an upper limit on the amount of extinction correction applied to a given star, 
as described more fully in \S\ref{sec:colorrangeapp}.

(ii) One important caveat discussed by \citet[][their \S2.1]{Majewski_11_RJCE} is that the RJCE method systematically 
overestimates the reddening to very late-type dwarfs (i.e., late K or M types) and 
stars with circumstellar shells or disks (e.g., asymptotic giant branch stars and pre-main sequence objects),
because their colors, even at near- and mid-IR wavelengths, are significantly redder than the blackbody-like colors of typical normal giants of the same
spectral type.  In late dwarfs, this is due to the presence of atmospheric molecular bands, including TiO and H$_2$O.
(In any case, the color-temperature relation shown in Figure~\ref{fig:dered}b is specifically for giant stars and diverges substantially
from that of dwarf stars around $T_{\rm eff} \lesssim 4000$ K, 
so the high fraction of ``mis-corrected'' stars with $\log{g} \gtrsim 4.5$ is, by definition, not surprising.)

The effect of this overestimation is that these stars are
systematically overcorrected to improperly blue colors.  
However, as also pointed out by \citet[][]{Majewski_11_RJCE}, 
the volume probed by M dwarfs within APOGEE's magnitude limits
is extremely small, so we do not anticipate many to fall in our sample,
and \S\ref{sec:colorrange} contains a description of the small
fraction ($\sim$1\%) of AGB stars anticipated in the APOGEE sample.
Furthermore, we note that this overcorrection may actually have improved our giant selection efficiency by removing some
cool dwarfs from the APOGEE color-magnitude selection box; this phenomenon is particularly helpful in the halo fields, 
which have an intrinsically higher dwarf/giant ratio.  
For this reason, we have chosen to continue using the RJCE dereddening method, 
cognizant of the fact that the corrected ``$(J-K_s)_0$'' values 
may not be an accurate representation of the intrinsic near-IR colors of these particular stars,
though this effect will be modulated by the inclusion of the SFD reddening values as upper limits on the stellar reddening corrections.

(iii) A uniform extinction law was assumed to convert the RJCE reddening into $E(J-K_s)$ across all fields, which may 
induce unaccounted-for systematic offsets if a field (or subset of stars in a field) has in reality 
a different relationship between $E(H-[4.5\mu])$ and $A[K_s]$.  
However, even assuming the most extreme level of variation in the sample (e.g., half the stars behind very dense ``dark cores''),
the induced scatter is on the order of $\lesssim$0.1 mag, and the observed stellar colors do not indicate that any 
significant fraction of stars lie in these extreme environments.  Therefore, we conclude that these possible variations
are not a major contributor to the observed scatter from the isochrone color-temperature relation
\citep[for discussions on variations in the NIR--MIR extinction law throughout a range of reddening environments, 
see e.g.,][]{Nishi_06_extlaw,Zasowski_09_extlaw,Gao_2009_MIRextlaw}.

\section{Stellar Clusters} \label{sec:clusters}

A large number of known stellar clusters, both open and globular, fall within the APOGEE survey footprint,
and we target these under two general classifications: ``calibration'' and ``science''.  
Calibration clusters (\S\ref{sec:calibclusters}) are defined as those with confirmed members having
well-determined stellar parameters and abundances.  The classification name is perhaps
something of a misnomer, since we expect to extract interesting science from these objects as well,
but it was chosen to distinguish them from the comparatively poorly-studied ``science'' clusters (\S\ref{sec:openclusters})
that lack well-characterized stellar data and definitive membership.

\subsection{Calibration Clusters} \label{sec:calibclusters}
Observations of cluster populations with well-characterized stellar parameters and abundances from existing high-resolution 
optical spectra are critical for testing and calibrating the
ASPCAP pipeline.  This step is essential for obtaining accurate abundances
of a large sample of widely distributed field giants, an integral part of the
survey goals listed in \S\ref{sec:intro}, and clusters are ideal calibrator resources
because they increase observing efficiency (since many targets can be observed simultaneously),
span a range of $\log g$ at a common abundance, 
and have a much larger quantity of published data than typical field stars.
(Most of the non-cluster calibrator sources are described in \S\ref{sec:gc}.) 

Furthermore, APOGEE is targeting additional
cluster members currently lacking such detailed abundances to improve our limited
understanding of globular cluster formation.  
APOGEE's access to the $H$-band enables it to measure
abundances for many of the light elements that show variations in globular
clusters, including C, N, O, Na, Mg, and Al.  Indeed, a ubiquitous feature of globular clusters is the suite of strong anti-correlations between the
relative abundances of light elements, such as Na--O, C--N, and Mg--Al
\citep[e.g,][]{Kraft_1994_GCabundancediffs,Shetrone_1996_AlMgEu-in-GCs}.
APOGEE's multi-object capability allows
for observations of large numbers of cluster stars, whose abundances will be
determined homogeneously.

Twenty-five of APOGEE's halo fields are placed deliberately on open and globular stellar
clusters with at least some stars having well-measured stellar parameters (27 clusters in total; see Table~\ref{tab:calibclusters}).  
The halo globular cluster fields present unique target selection challenges due to
the highly variable target densities in these fields, so we have developed a
special targeting algorithm and prioritization scheme for these cases.  
One of the main challenges for target selection in the globular clusters themselves is
avoiding fiber collisions among closely packed cluster members.
We take advantage of the multiple designs made for a given field 
to carefully assign stars to designs in which they will not collide with their neighbors
(which are assigned to different designs), thus minimizing the loss of these valuable targets to avoidable fiber collisions.

\begin{deluxetable*}{llccl} \tablewidth{0 pt}
\tablecaption{Calibration Clusters}
\tablehead{ \colhead{Cluster Name} & \colhead{Alt Name} & \colhead{[Fe/H] (ref.)} & \colhead{log(age/yr)\tablenotemark{a} (ref.)} & \colhead{APOGEE Field}}
\tabletypesize{\footnotesize}
\startdata
Berkeley 29 & & $-0.31\pm0.03$ (1) & 9.0 & {\it 198+08} \\ 
Hyades & & $+0.13\pm0.01$ (2) & 8.8 (17) & {\it HYADES} \\
M45 & Pleiades & $+0.03\pm0.02$ (3) & 8.1 & {\it PLEIADES} \\
NGC 188 & & $-0.03\pm0.04$ (4) & 9.6 & {\it N188} \\
NGC 2158 & & $-0.28\pm0.05$ (4) & 9.3 (18) & {\it M35N2158} \\
NGC 2168 & M35 & $-0.21\pm0.10$ (5) & 8.0 & {\it M35N2158} \\
NGC 2243 & & $-0.48\pm0.15$ (6) & 9.6 (19) & {\it N2243} \\
NGC 2420 & & $-0.20\pm0.06$ (4) & 9.0 & {\it N2420} \\
NGC 2682 & M67 & $-0.01\pm0.05$ (4) & 9.4 & {\it M67} \\
NGC 4147 & & $-1.78\pm0.08$ (7) & GC & {\it N4147} \\
NGC 5024 & M53 & $-2.06\pm0.09$ (7) & GC & {\it M53} \\
NGC 5272 & M3 & $-1.50\pm0.05$ (7) & GC & {\it M3} \\
NGC 5466 & & $-2.31\pm0.09$ (7) & GC & {\it N5466} \\
NGC 5634 & & $-1.88\pm0.13$ (8, 9) & GC & {\it N5634SGR2} \\
NGC 5904 & M5 & $-1.33\pm0.02$ (7) & GC & {\it M5PAL5} \\
NGC 6171 & M107 & $-1.03\pm0.02$ (7) & GC & {\it M107} \\
NGC 6205 & M13 & $-1.58\pm0.04$ (7) & GC & {\it M13} \\
NGC 6229 & & $-1.43$ (8, 10, 11) & GC & {\it N6229} \\
NGC 6341 & M92 & $-2.35\pm0.05$ (7) & GC & {\it M92} \\
NGC 6715 & M54 & $-1.49\pm0.02$ (8, 12) & GC & {\it M54SGRC1} \\
NGC 6791 & & $+0.47\pm0.07$ (13) & 9.6 & {\it N6791} \\
NGC 6819 & & $+0.09\pm0.03$ (14) & 9.2 & {\it N6819} \\
NGC 6838 & M71 & $-0.82\pm0.02$ (7) & GC & {\it M71} \\
NGC 7078 & M15 & $-2.33\pm0.02$ (7) & GC & {\it M15} \\
NGC 7089 & M2 & $-1.66\pm0.07$ (7) & GC  & {\it M2} \\
NGC 7789 & & $+0.02\pm0.04$ (4) & 9.2 & {\it N7789} \\
Palomar 5 & & $-1.41\pm0.20$ (8, 15, 16) & GC & {\it M5PAL5}
\enddata
\tablenotetext{a}{A ``GC'' denotes a globular cluster \citep{Harris_1996_MWGCs,Harris_2010_newMWGCs}.  
Open cluster ages without a reference are drawn from WEBDA: \url{http://www.univie.ac.at/webda/}.}
\tablerefs{(1) \citet{Sestito_2008_OCabundances}, (2) \citet{Paulson_2003_hyades}, (3) \citet{Soderblom_2009_pleiades}, 
(4) \citet{Jacobson_2011_OCabundances}, (5) \citet{Barrado_2001_m35}, (6) \citet{Gratton_1994_ngc2243}, 
(7) \citet{Carretta_2009_GCmetals}, (8) \citet{Harris_2010_newMWGCs}, (9) \citet{Zinn_1984_GCs},
(10) \citet{Searle_1978_haloclusters}, (11) \citet{Wachter_1998_GCmetals}, (12) \citet{Carretta_2010_m54sgrOCen},
(13) \citet{Carretta_2007_metalrichOCs}, (14) \citet{Bragaglia_2001_ngc6819}, (15) \citet{Geisler_1997_washphotGCs}, 
(16) \citet{Smith_2002_pal5}, (17) \citet{Perryman_1998_hyades}, (18) \citet{Carraro_2002_ngc2158}, (19) \citet{Houdashelt_1992_OCgiants}}
\tablecomments{Metallicities from the \citet{Harris_2010_newMWGCs} catalog (ref.\,8) have been shifted 
from their original values onto the scale of \citet{Carretta_2009_GCmetals}; 
the quoted uncertainties are drawn from the original references cited and are therefore approximate.}
\label{tab:calibclusters}
\end{deluxetable*}

To further increase our yield of cluster members, we also include ``faint''
targets: stars that are $\leq$0.8 magnitudes fainter than the magnitude limit of the
longest cohort in the field.  
Even though the ``faint'' targets will not meet the magnitude limit required to reach ${\rm S/N} =
100$ for the number of visits in a given field, 
we typically expect the ``faint'' targets to have ${\rm S/N} \sim 70$, a level
sufficient for measuring the stellar parameters and abundances of some elements.
This strategy allows us to probe to fainter magnitudes without investing twice
the number of visits for a given field to increase the S/N from $\sim$70 to 100
for only a few stars.  ``Faint'' targets have bit \atone = 29 set.

Unlike the densely-packed clusters, 
the halo fields outside of the cluster boundaries have a very low density of stars, 
necessitating the targeting of stars which would otherwise be avoided.  
For example, fibers remaining after all possible cluster members and Washington+DDO51 giants have been assigned
may be placed on stars lacking Washington+DDO51 classification or even on Washington+DDO51 dwarfs.
In the cluster halo fields and other fields with Washington+DDO51 photometry, 
the fibers available for these lower priority stars are generally sufficiently few in number 
that we do not attempt to sample the stars evenly in magnitude, like the $N^{\rm th}$
sampling scheme for the ``standard'' target selection described in \S\ref{sec:magsample};
instead, these targets are chosen (if there are enough to present a choice) at random with respect to magnitude.
(For the halo fields without Washington+DDO51 data, the stars are sampled using the method described for the short cohorts in \S\ref{sec:magsample}.)
In addition, for most halo fields, we widen the color range relative to the ``standard'' target selection ($[J-K_s]_0 \geq 0.3$, rather than 0.5;
Appendix~\ref{sec:globclustercmrange}) to increase the number of potential targets, especially in the
higher priority classes described below.

In contrast to the selection of field RG stars, the importance of particular
stars in the halo fields varies wildly, from the invaluable cluster members to
the low priority field dwarfs.  To ensure that we are targeting the most
useful stars first, the halo globular cluster target selection algorithm
separates stars into the following priority classes based on their desirability:
\begin{enumerate}
\item cluster members confirmed via high resolution optical spectroscopy,
  with existing measurements of stellar parameters and abundances
  (ranging from a few stars to $>$50 stars in the most well-studied clusters),
\item cluster members confirmed via proper motions (proper motion-based membership probability $\geq$90\%) 
\item cluster members confirmed via radial velocity (RV) measurements,
\item likely cluster members with proper motion membership probability $50-90$\%,
\item ``faint'' cluster members confirmed via high resolution optical
  spectroscopy,
\item ``faint'' cluster members confirmed via proper motions (probability
  $\geq$90\%),
\item ``faint'' cluster members confirmed via RV measurements,
\item ``faint'' likely cluster members with proper motion membership probability $50-90$\%,
\item SEGUE overlap targets (\S\ref{sec:segue}), 
\item ``faint'' SEGUE overlap targets, 
\item Washington+DDO51 giants (\S\ref{sec:halo}) with $(J-K_s)_0 \geq 0.3$,
\item ``faint'' Washington+DDO51 giants with $(J-K_s)_0 \geq 0.3$,
\item red ($[J-K_s]_0 \geq 0.5$) targets without Washington+DDO51 giant/dwarf
  classification,
\item blue ($0.3 \leq [J-K_s]_0 < 0.5$) targets without Washington+DDO51 giant/dwarf
  classification, and
\item Washington+DDO51 dwarfs with $(J-K_s)_0 \geq 0.3$.
\end{enumerate}

This prioritization scheme is absolute in the sense that stars of a higher
priority class will always be selected over a star from a lower priority class
within the same cohort
(exceptions are noted in Appendix~\ref{sec:globclusterexceptions}).  
Stars belonging to multiple priority
classes are submitted as members of their highest class but may have multiple targeting bitmasks set from the multiple classes.
The cluster members are sufficiently valuable that we ignore any 2MASS quality flags for
these stars.  
The cluster member lists, along with the membership probabilities based on spectroscopy, RVs, and proper motions,
will be presented in subsequent papers analyzing the APOGEE data.
All of the calibration cluster member stars have bits \attwo = 2
and/or 10 set, as appropriate, Washington+DDO51 giants or dwarfs have bit \atone = 7 or 8 set, respectively, 
and SEGUE overlap targets have bit \atone = 30 set.

We utilize the same target selection algorithm for the few long
non-cluster halo fields,
with the obvious exception of the cluster member classes; 
the prioritization for these fields
starts with any SEGUE overlap targets that are present. 

\subsection{Open Clusters \& Cluster Candidates} \label{sec:openclusters}
In addition to the well-studied stellar clusters in Table~\ref{tab:calibclusters}, 
APOGEE is targeting a large number of open clusters
and cluster candidates in the disk that either have no 
previous measurements of cluster parameters (i.e., age, distance, [Fe/H]) 
or only have measurements based on a very small number of stars.  
The detection and study of open clusters have proven to be of great benefit to 
understanding the chemo-dynamics of the Galactic disk \citep[e.g.,][]{Friel_1995_OCreview}.
Open clusters represent single stellar populations at a single distance,
with a common chemical composition and a common age. 
But unlike globular clusters, the open clusters as a population span the range of ages necessary
to trace the recent history of star formation in the disk.
With comparisons to theoretical stellar isochrones, the cluster's age, distance,
and metallicity can be estimated from photometry alone.  

In addition, {\it spectroscopic} datasets of open cluster member stars provide
(i) a precise estimate of the mean kinematics for the cluster, 
(ii) independent estimates of the mean chemical abundances of the system, 
and
(iii) RV membership discrimination, which tightens the constraints on points (i) and (ii).
Furthermore, spectroscopically-derived chemistry can be combined with stellar photometry 
to break the age-metallicity-distance degeneracies inherent in isochrone-fitting 
and to determine robust ages for the stellar population.
The combination of ages and chemical compositions for multiple clusters provides a direct
assessment of the chemical evolution of the Galactic disk over the past several Gyr.
The open cluster population thereby provides a rich dataset through which 
 the star-formation history and chemical evolution of the Galaxy can be explored 
 and compared to the predictions from numerical chemodynamical models.

APOGEE's wavelength coverage allows it to probe heavily reddened clusters previously inaccessible to many spectroscopic surveys,
and the RV precision is sufficient to distinguish the kinematical signature of cluster member candidates from that of the underlying stellar disk.
Furthermore, the spectral resolution of APOGEE enables measurement of both [Fe/H] and [$\alpha$/Fe]
with the precision required for full chemo-dynamical analysis of the cluster member stars. 

\begin{figure*} 
\begin{center}
\includegraphics[angle=90,width=0.99\textwidth]{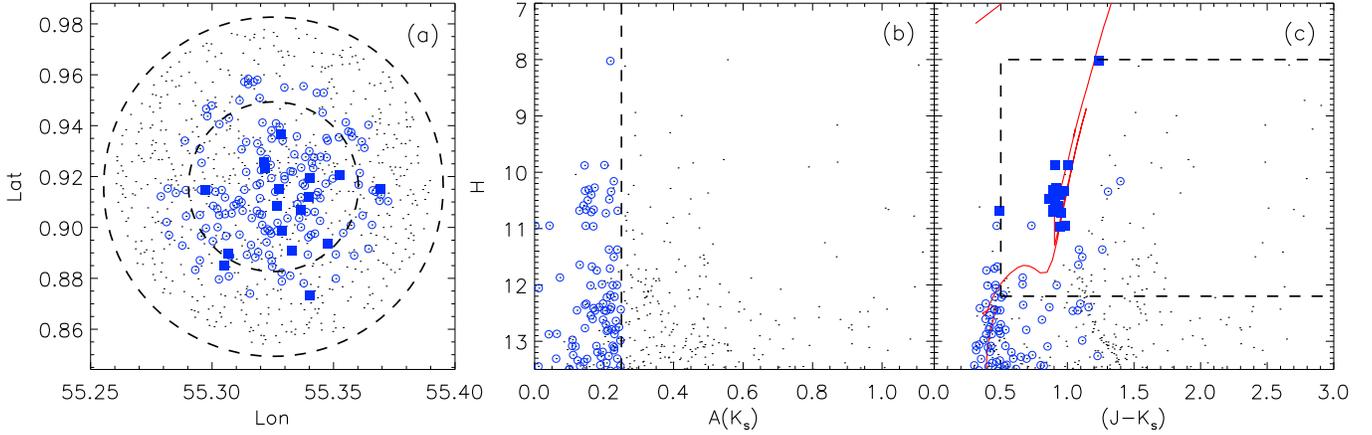} 
\end{center}
\caption{
Selection of candidate open cluster members, using NGC~6802 as an example.
{\it (a):} Spatial distribution of stars within 2 cluster radii ($R$) of the cluster center
(inner dashed ring: $R = 2$ arcmin, outer dashed ring: 2$R = 4$ arcmin).
All stars are shown as black points, open blue circles are overplotted on the stars within 1.2$R$ that
meet the extinction criteria shown in {\it (b)}, 
and filled blue squares are overplotted on the final high-priority targets.
{\it (b):} Distribution of $A(K_s)$ extinction values for all stars within 2$R$ (black points).  
Open blue circles are overplotted on those stars within 1.2$R$ and with $0 \leq A(K_s) \leq 0.25$.
{\it (c):} 2MASS CMD of all stars within 2$R$ (black points).  Open blue circles are on the stars defined in {\it (b)},
and the dashed lines enclose the color-magnitude zone used to select the final target sample,
indicated by solid blue squares.
An 0.7 Gyr isochrone \citep{Girardi_02_isochrones}
with distance modulus $\mu = 12.5$  and reddening $E(J-K_s)=0.35$ is shown in red. 
}
\label{fig:opencluster}
\end{figure*}

The $\sim$450 open clusters and cluster candidates falling serendipitously within APOGEE's fields 
were identified in the online catalog based on \citet{Dias_2002_OCcatalog},
and the majority have no spectroscopically confirmed members.
Lists of candidate members were generated using the methodology to be
explained in detail by Frinchaboy et al. {\it in prep}.  
Figure~\ref{fig:opencluster} demonstrates the stages of this identification.
In summary, this method isolates overdensities in both the spatial and extinction distributions of stars 
within a few cluster radii of the published cluster center
\citep[with stellar extinctions derived using the RJCE technique;][]{Majewski_11_RJCE}.  
Then, $(J-K_s)$ color and $H$ magnitude limits, as well as visual examination of the corresponding NIR CMD, 
are used to identify the highest-priority cluster stars, particularly those most likely to be cluster RC or RGB stars.
Though the open clusters have significantly lower stellar spatial density 
than many of the calibration clusters (\S\ref{sec:calibclusters}),
we still minimize fiber collisions by distributing targets among multiple designs for the cluster's field,
to maximize the final number of cluster targets observed.
Open cluster candidate members targeted with this method have bit \atone = 9 set.

\section{Additional Targets} \label{sec:addltargs}
The wavelength coverage of APOGEE, 
combined with the observational cadence and 
multi-fiber capability, enables a wide range of science beyond the primary survey goals.  
A number of secondary target classes have been defined to meet additional science goals,
which either were included in the original survey scheme or have arisen from the community as APOGEE's capabilities became better understood.
Here, we present a brief overview of the customized targets found in the APOGEE sample.

\subsection{Special Galactic Bulge Giants and Supergiants} \label{sec:gc}
Extensive photometric and spectroscopic studies have targeted Baade's Window, 
a low-extinction region of the bulge at $(l,b) \sim (1^\circ, -4^\circ)$, and other bright bulge stars
with lower-than-average extinction.
These efforts have resulted in a moderate sample of bulge stars, especially in Baade's Window, 
with well-determined parameters and chemical abundances from medium- and high-resolution spectroscopy.
APOGEE is also targeting many of these stars to calibrate the ASPCAP pipeline, confirm
and expand upon the interesting abundance patterns observed in some of the stars,
and probe the abundances and kinematics of the bulge. 
Our sample includes late-type bulge giants from 
\citet{Cunha_06_bulgespectra}, \citet{Fulbright_2006_baadewinchem}, 
\citet{Zoccali_06_oxygeninbulge}, \citet{Lecureur_07_bulgechemform},
\citet{Fulbright_07_alphainbulge}, \citet{Cunha_08_bulgeflourine}, 
\citet{Melendez_08_bulgechemistry},
\citet{Ryde_09_bulgespectra}, \citet{Ryde_10_bulgespectra}, 
and \citet{Alves-Brito_2010_bulgethickdisk},
all observed with high resolution ($R \gtrsim 30,000$).
We also include targets from earlier studies at lower resolution ($15,000 < R < 30,000$),
namely the work of \citet{McWilliam_94_KgiantsBW}, \citet{Rich_05_MgiantsBW},
and \citet{Zoccali_08_bulgechem}.

There is considerable overlap among these studies --- for example, all of the stars
in the samples of \citet{Melendez_08_bulgechemistry}, \citet{Ryde_09_bulgespectra},
and \citet{Alves-Brito_2010_bulgethickdisk} were selected from 
the stars observed earlier by \citet{Fulbright_2006_baadewinchem}.
Since different wavelength ranges, instruments, and techniques were
used when analyzing these overlapping samples, they provide an excellent measure
of the systematic scatter between studies.  Coordinates and 2MASS names for the
stars identified only by their \citet{Arp_1965_baadewinchart} numbers are drawn
from Table~1 of \citet{Church_2011_arpbaadewin}.  
 
In addition to the Baade's Window studies, APOGEE has a number of targets in common
with the BRAVA \citep[][]{Rich_2007_brava,Kunder_2012_bravaDR} 
and ARGOS \citep{Ness_2012_argos} surveys.
Both of these programs were carried out from the Southern hemisphere, 
while APOGEE operates from the North, but the surveys complement each other in the bulge.
These targets --- $\sim$175 from ARGOS and $\sim$90 from BRAVA ---
are located in APOGEE's {\it BAADEWIN} and {\it BRAVAFREE} fields.

In a total of eight bulge and inner midplane fields with $6^\circ \lesssim l \lesssim 20^\circ$, 
we target $\sim$110 giants and supergiants with existing
{\it HK}-band low-resolution spectroscopy from \citet{Comeron_2004_innerSGs}.
This subsample is drawn from the larger Comer{\'o}n et al.\,spectroscopic catalog,
which consists of stars with $K < 10$, with strong CO bands, and without contamination 
from detector artifacts or nearby bright sources, as assessed from visual inspection of NIR imaging
\citep[see details in][]{Comeron_2004_innerSGs}.
In the {\it GALCEN} field, spanning the Galactic Center, 
we also include M supergiants from the Quintuplet and Central clusters with abundances measured from high resolution spectroscopy
\citep{Carr_2000_supergiantIRS7,Ramirez_2000_GalCenFEH,Cunha_2007_galcenspectra,Davies_2009_galcensupergiants},
other giants and supergiants selected by low resolution $K$-band spectroscopy
\citep[][and M.~Schultheis, T.~R.~Geballe, and C.~DeWitt, private communication]{Schultheis_03_bulgespectra,Blum_2003_galcenstars,Mauerhan_2007_galcenSGs,Mauerhan_2008_galcenXrays,Liermann_2009_quintupletcluster},
and probable supergiants, as identified by their {\it JH$K_s$} photometry, within $0.25^\circ$ of the Galactic midplane.

In two fields ({\it 358+00} and {\it GALCEN}), 
we specifically target $\sim$38 AGB candidates from the sample of \citet{Schultheis_03_bulgespectra}, 
who used H$_2$O and CO absorption, ISOGAL mid-IR excesses, and light-curves
to identify these stars based on their high mass-loss rates, chemical composition, and/or variability.  
These targets have $7 \leq H \leq 12$, a range chosen to assure high continuum S/N 
but also to avoid saturation, given the potential $1-2$ mag $H$-band variability.
The final sample was selected and prioritized based on the presence of extant photometry (spanning $1.2-15$ $\mu$m), 
low-resolution near-IR spectra \citep[{\it JHK} bands;][]{Schultheis_03_bulgespectra},
mid- to far-IR spectra (e.g., from ISO or {\it Spitzer}), and/or high resolution optical spectra 
with characterization of Li and s-process elements),
all of which make these valuable sources for cross-calibration of stellar parameters and abundances.
These AGB targets have among the coolest effective temperatures ($T_{\rm eff} \lesssim 3500$ K) in the APOGEE stellar sample,
which makes them useful for developing and testing the ASPCAP pipeline at the low end of the temperature grid 
\citep[Garcia Perez et al., {\it in prep.};][]{Meszaros_2012_APOGEEatmogrids}.
Included in this sample are some of the intrinsically brightest AGB stars in the bulge (e.g., those with $M_{\rm bol} < -7$),
which may comprise the most massive AGB stars \citep[see e.g.,][]{GarciaHernandez_2006_RbAGB,GarciaHernandez_2009_MagClAGB}.

Furthermore, this bulge AGB sample provides the opportunity to study AGB nucleosynthesis in a relatively high metallicity environment. 
For example, the $^{12}{\rm C}/^{13}{\rm C}$ isotopic ratio, which is not easily accessible at optical wavelengths in these O-rich stars,
can be derived from molecular features in the $H$-band APOGEE spectra \citep{Smith_2012_FTSredgiants}
and used as an indicator of ``hot bottom burning''  \citep[e.g.,][]{GarciaHernandez_2007_LiZrAGB}.
Ca, Na, and Al are other important elements accessible by APOGEE
that are expected to be strongly altered by the nucleosynthetic processes experienced by AGB stars
\citep[e.g.,][and references therein]{Karakas_2012_AGBheavyelement}.
Finally, in contrast to ordinary M giants, AGB stars have a complex atmospheric structure that poses large challenges to reliable modeling of their atmospheres. 
Given their large pulsation amplitudes, AGB stellar atmospheres can only be described by 
advanced hydrodynamical model atmospheres that are coupled with dust formation \citep[e.g.,][]{Hofner_2012_AGBdustgrains}.
The sample of AGB stars for APOGEE will provide important inputs for developing and testing these complex model atmospheres.  

In the targeting bitmasks for all of these stars, \attwo = 11 indicates a spectroscopically-confirmed giant, \attwo = 12 indicates a supergiant, and 
\attwo = 2 indicates the star is also a stellar parameter and/or abundances standard.

\subsection{Sagittarius dSph Core and Tails} \label{sec:sgr}
The Sagittarius (Sgr) dwarf spheroidal (dSph) galaxy is one of the most massive of the Milky Way's surviving satellite galaxies (exceeded only by 
the Magellanic Clouds), and its distorted core and extensive tidal tails offer an opportunity to observe the ongoing process of ``galactic cannibalism'' that
has been so influential in assembling our present-day Galaxy \citep[e.g.,][]{Ibata_94_SgrdSph,Newberg_02_monocerosring,Majewski_03_SgrdSph}.  
The core of Sgr lies $\sim$30 kpc from the Sun
in the direction of the Galactic Center, with a latitude $\sim$12$^\circ$ below the midplane, 
and the tails have been observed to wrap more than 360$^\circ$ around the Milky Way.  
The properties of these components --- including kinematics and metallicity gradients 
\citep{Chou_07_SgrStreamFeH,Keller_10_SgrArmChem,Chou_10_SgrStreamAbun} --- record the history
of not only the Sgr dSph itself but also its interaction with the Milky Way.

APOGEE has placed fields covering the Sgr core and multiple locations along the tidal streams.  
In the stream fields, candidate Sgr members are targeted based on a selection of 2MASS M giants \citep{Majewski_03_SgrdSph}.  In the core, the same
M giant selection process is used, supplemented by targets with membership based on 
kinematics derived from medium resolution spectroscopy \citep[$R \sim 15,000$;][]{Frinchaboy_2012_Sgrcorekinematics}.
The target bitmasks of these stars have bit \atone = 26 set.

\subsection{{\it Kepler} Targets} \label{sec:kepler}
Since its launch in 2009, NASA's {\it Kepler} satellite has been monitoring the photometric variations of 
$\sim$$1.5 \times 10^5$ stars in a 105 deg$^2$ patch of  sky towards the constellation Cygnus \citep{Borucki_2010_keplerintro}.
{\it Kepler}'s high temporal resolution and photometric precision have led not only to the discovery of many extra-solar transiting planets 
($>$115 confirmed planets and thousands of candidates, as of April 2013)
but also to the further development of asteroseismology (the study of stellar pulsations) 
as a unique and powerful probe of stellar interiors in large samples of stars. 
Through the analysis of solar-like oscillations, fundamental stellar parameters, 
such as mass and radius (and thus $\log{g}$), 
can be determined with high accuracy \citep[e.g.,][]{Stello_2009_asteroseismology,Gilliland_10_KeplerAsteroseis}. 
In addition to providing calibration data for ASPCAP,
the combination of these parameters with APOGEE's high-precision chemical abundances enables measurements of stellar
{\it ages} to $\lesssim20$\% accuracy. 
Ages are notoriously difficult quantities to obtain for field (i.e., non-cluster) stars,
but they are crucial for probing such Galactic properties as the star formation history, radial mixing efficiency, and
age-metallicity relation.

To that end, APOGEE has adopted a targeting plan to tile the {\it Kepler} field. 
By fortuitous coincidence, the size of a single $\sim$7~deg$^2$ APOGEE field is well matched to 
that of a single {\it Kepler} CCD module projected on the sky, and
because the magnitude distribution of red giants in the {\it Kepler} field is dominated by stars within the range $7<H<11$, 
a one-hour visit is sufficient to achieve APOGEE's goal of ${\rm S/N} = 100$ for the majority of {\it Kepler} giants. 
The 21 ``APOGEE-{\it Kepler}'' fields, each centered on one of {\it Kepler's} 21 module centers,
were allocated 40 visits in total (approximately 40 hours of integration), 
with unique targets selected for each visit to a field. 

By combining fields that APOGEE had already planned to observe during Year 1 (e.g., those containing the open clusters NGC~6791 and NGC~6819)
with the 40 additional visits to APOGEE-{\it Kepler} fields over the course of the survey, 
the plan is to target $\sim$10,000 stars within the {\it Kepler} survey footprint. 
The final count will be dependent on weather and fiber collisions. 
The spatial density of potential targets across the {\it Kepler} field of view determines the distribution of the 
40 allotted visits among the 21 APOGEE-{\it Kepler} fields, though each field is guaranteed at least one visit. 
All {\it Kepler} targets in the APOGEE-{\it Kepler} fields are required to lie within the magnitude range 
$7 \leq H \leq 11$ and have an effective temperature cooler than 6500 K, as determined by \citet{Pinsonneault_2012_KICTeffrevision}.

Stars with detected solar-like oscillations comprise the majority ($\sim$87\%) of the sample.
These include $\sim$640 dwarfs observed with {\it Kepler}'s ``short cadence'' mode,
with the remaining spots reserved for $\sim$8000 red giant targets with longer-period oscillations. 
Because the number of available candidates for these spots is much larger 
\citep[the {\it Kepler} dataset contains $\sim$16,000 red giants; e.g.,][]{Hekker_2011_RGsinKepler},
a number of important subsets were prioritized first, such as
the sample of giants identified by the {\it Kepler} Asteroseismic Consortium as having high S/N detections. 
Other valuable populations include members of the {\it Kepler} open clusters NGC~6791, NGC~6819, NGC~6811, and NGC~6866, 
all of which have targeting bit \atone = 9 set. 
Probable members of the Galactic halo were selected using Washington photometry, proper motions, and low resolution ($R \sim 2000$) 
spectra from MARVELS/SEGUE target pre-selection data.  
Rapid rotators and stars with unusual seismic properties were also identified and included in the higher priority groups. 
The remaining targets in the asteroseismic sample were prioritized based on a number of factors, 
including brightness and length of the {\it Kepler} observation baseline. 
If the evolutionary state was known from seismic measurements of mixed modes \citep[e.g.,][]{Bedding_2011_RCvsRGB}, 
first ascent RGB stars were favored over RC stars to avoid complications due to mass loss on the upper RGB.

A smaller fraction ($\sim$13\%) of the sample was reserved for other interesting stars in the {\it Kepler} field,
the bulk of which comprise a distinct ancillary program targeting $\sim$1200 cool dwarfs, 
as described in Appendix~\ref{sec:kepcooldwarfs}.
The remaining APOGEE-{\it Kepler} targets include all 
of {\it Kepler}'s planet-host candidates identified in \citet{Batalha_2012_keplerplanets} 
that meet the $H$ mag and T$_{\rm eff}$ restrictions given above, 
along with $\sim$38 eclipsing binary stars identified during the asteroseismic analysis (see also Appendix~\ref{sec:EBs}).

All targets with asteroseismic detections have targeting bit \atone = 27 set,
and the planet-host candidates have targeting bit \atone = 28 set,
In addition, {\it UBV} and {\it griz} photometry acquired for many of the {\it Kepler} targets 
\citep{Brown_2011_KIC,Everett_2012_keplerUBV}
will be provided where available.

\subsection{SEGUE Overlap Targets} \label{sec:segue}
The Sloan Extension for Galactic Understanding and Exploration \citep[SEGUE;][]{Yanny_09_SEGUE}, one of the SDSS-II programs,
obtained {\it ugriz} imaging and medium-resolution optical spectroscopy ($R \sim 1800$) for $\sim$240,000 targets, which span almost all 
stellar types and populations and reside primarily in the Milky Way halo.  
SEGUE-2 
continued this approach, observing $\sim$120,000 stars during the first year of SDSS-III operations (Rockosi et al.,\ {\it in prep}).
The SEGUE Stellar Parameters Pipeline \citep[SSPP;][]{Lee_08_SSPP1,Lee_08_SSPP2,AllendePrieto_08_SSPP3} 
has calculated effective temperatures, surface gravities, and metallicities for the spectra in both surveys.

Overlap between the SEGUE-1 and -2 pointings and the APOGEE halo fields provides 
SEGUE/SSPP targets for inclusion in the APOGEE sample, 
increasing the number of stars with which to test ASPCAP and inform its development.
Approximately $\sim$1000 SEGUE targets were targeted at least once by APOGEE in Year 1, with
further visits to these and $\sim$100 additional stars anticipated in Years 2--3.  These targets were selected based 
solely on their location within the APOGEE FOV and their $H$ magnitudes, and all have bit \atone = 30 set.

\subsection{Ancillary Programs} \label{sec:anc1}
In keeping with SDSS tradition, the APOGEE team announced calls for ancillary science proposals ---
one before commissioning operations began, and a second one halfway through Year 1 observations.
In total, $\sim$5\% of APOGEE's fiber-visits are allotted to ancillary science programs, 
which range in scope from a single 1-hour integration on a particular star to 
multiple designs with all science fibers in each dedicated to the ancillary program. 
All of these data 
will be available along with the other APOGEE spectra in the data releases according to the timeline of their observation.  

Descriptions of the science goals and target selection processes for each ancillary program may be found in Appendix~\ref{sec:anc}.
In addition to their individual program bitmask flags (Table~\ref{tab:targflags}), all ancillary targets have bit \atone = 17 set.

\section{Targeting Information To Be Included In Data Releases} \label{sec:release}
The first public release of APOGEE data is in the summer of 2013, as part of the SDSS-III Data Release 10 (DR10).
This release comprises data --- reduced spectra, RVs, best-fit synthetic spectra,
basic stellar parameters ($T_{\rm eff}$, $\log{g}$, [Fe/H], microturbulence), 
and abundances ([C/Fe], [N/Fe], total [$\alpha$/Fe]) ---
from the first year of survey operations, spanning September 2011 to July 2012.  

In addition, where available, supplementary information on each target
will be provided to help users account for any relevant
observational biases present in their queried dataset, including: 
\begin{itemize}
\item the photometry and photometric uncertainties 
used in the target selection, from 2MASS, {\it Spitzer} IRAC, WISE, and our own $M$, $T_2$, and DDO51 photometry;
\item WISE photometry from the first all-sky data release, as well as any pre-release data used in the target selection;
\item {\it UBVgriz} photometry and stellar parameters derived from asteroseismological measurements for the APOGEE-{\it Kepler} sample;
\item RJCE and SFD reddening and extinction values (those used in the target selection and other values); 
\item the dwarf/giant classifications for stars with Washington+DDO51 data; and
\item proper motions, which were collected to provide corrected coordinates 
on the drilled plates but {\it not} used in the selection of normal APOGEE targets.
\end{itemize}
A subset of this supplementary information will also be provided for non-targeted 2MASS sources in the APOGEE fields.

Furthermore, DR10 will contain tables of APOGEE's fields, designs, and plates 
(identified by their respective location, design, and plate IDs)
along with their useful information: 
\begin{itemize}
\item {\it fields:} central coordinates, number of visits, and classification (e.g., disk, bulge, calibration cluster);
\item {\it designs:} angular radius, short/medium/long cohort versions, short/medium/long fiber allocations, 
and cohort magnitude ranges; and
\item {\it plates:} hour angle, temperature, and observation epoch for which each was drilled.
\end{itemize}

We anticipate annual releases of APOGEE data with each SDSS-III Data Release,
which will also include all previously released data, improved with updated software or analysis
where possible, to produce a homogeneous set of data.

\begin{acknowledgments}

We thank Michael Blanton, Jo Bovy, Nathan De Lee, Jim Gunn, Inese Ivans, Sarah Martell, Demitri Muna,
and Young Sun Lee for useful discussions and assistance with documentation and targeting, and
we are very grateful to the WISE team for graciously permitting us access to their
pre-release photometric data.
GZ has been supported by a NASA Earth \& Space Science Fellowship and 
an NSF Astronomy \& Astrophysics Postdoctoral Fellowship under Award No.\ AST-1203017.
JAJ and PF were supported in part by the National Science
Foundation under Grant No.\ PHYS-1066293 and the hospitality of the Aspen
Center for Physics.
GZ and DF are grateful to the Astrophysical Research Consortium (ARC)
and the Sloan Digital Sky Survey (SDSS) for bestowing on them Young SDSS Astronomer Travel 
Assistance Awards to attend SDSS Collaboration Meetings.
DF also thanks the Spanish Ministerio de Ciencia e Innovaci\'on for supporting his research. 

Funding for SDSS-III has been provided by the Alfred P. Sloan Foundation, the Participating Institutions, the National Science Foundation, and the U.S. Department of Energy Office of Science. The SDSS-III web site is \url{http://www.sdss3.org/}.

SDSS-III is managed by the Astrophysical Research Consortium for the Participating Institutions of the SDSS-III Collaboration including the University of Arizona, the Brazilian Participation Group, Brookhaven National Laboratory, University of Cambridge, Carnegie Mellon University, University of Florida, the French Participation Group, the German Participation Group, Harvard University, the Instituto de Astrofisica de Canarias, the Michigan State/Notre Dame/JINA Participation Group, Johns Hopkins University, Lawrence Berkeley National Laboratory, Max Planck Institute for Astrophysics, Max Planck Institute for Extraterrestrial Physics, New Mexico State University, New York University, Ohio State University, Pennsylvania State University, University of Portsmouth, Princeton University, the Spanish Participation Group, University of Tokyo, University of Utah, Vanderbilt University, University of Virginia, University of Washington, and Yale University. 
This publication makes use of data products from the Two Micron All Sky Survey, which is a joint project of the University of Massachusetts and the Infrared Processing and Analysis Center/California Institute of Technology, funded by the National Aeronautics and Space Administration and the National Science Foundation,
and from the Wide-field Infrared Survey Explorer, which is a joint project of the University of California, Los Angeles, and the Jet Propulsion Laboratory/California Institute of Technology, funded by the National Aeronautics and Space Administration.
This work is also based in part on observations made with the {\it Spitzer Space Telescope}, which is operated by the Jet Propulsion Laboratory, California Institute of Technology under a contract with NASA.
We acknowledge use of NASA's Astrophysics Data System Bibliographic Services and of the VizieR catalogue access tool, CDS, Strasbourg, France.

\end{acknowledgments}

\bibliographystyle{aa}
\bibliography{/Users/GailZasowski/Documents/MW_Dust/reflib}

\begin{appendix}

\section{Glossary} \label{sec:glossary}
This Glossary contains SDSS- and APOGEE-specific terminology which will appear throughout the survey documentation and data releases,
particularly focusing on those terms likely to be ambiguous or unfamiliar to those unaccustomed to working with SDSS or APOGEE data.
\begin{description}
\item[Ancillary Target] Target observed as part of an approved ancillary program to utilize APOGEE's capabilities for interesting science beyond
the primary survey goals (\S\ref{sec:anc1} and Appendix~\ref{sec:anc}).
\item[ASPCAP] ``APOGEE Stellar Parameters and Chemical Abundances Pipeline''; the software pipeline used to calculate basic stellar parameters 
($T_{\rm eff}$, log~$g$, [Fe/H], [$\alpha$/H], $\xi$) and other elemental abundances (Garcia Perez et al., {\it in prep}).
\item[Cohort] Set of targets in the same field observed together for the same number of visits (\S\ref{sec:magrange}).  A given plate may have multiple cohorts on it.
\item[Design] Set of targets selected together for drilling on a plate or plates; may consist of up to one each of short, medium, and long cohorts.
A design is identified by an integer Design ID.  Changing a single target on a design results in a new design.
\item[Design ID] Unique integer assigned to each design.
\item[Fiber Collision] An attempt to place, on the same design, two targets separated by less than the diameter of the protective ferrules
around each fiber (the APOGEE ferrules are 71.5 arcsec in diameter).  The SDSS-III plate design software will assign only the higher-priority target to be drilled;
the lower-priority target is ``rejected''.
\item[Field] Location on the sky, defined by central coordinates and radius (\S\ref{sec:fieldplan}).  Fields can be identified by a 
string Field Name (e.g., {\it ``090+08''}) or integer Location ID (e.g., 4102).
\item[Location ID] Unique integer assigned to each field on the sky.
\item[Normal Targets] Science targets selected with the data quality criteria, color and magnitude criteria, and
magnitude sampling algorithms as defined in \S\ref{sec:targselection} of this paper.  Contrasted with ``special'' targets --- e.g., ancillary and calibration targets.
\item[Plate] Unique piece of aluminum with a design drilled on it.  Note that while ``plate'' is commonly used interchangeably with ``design'', 
multiple {\it plates} may exist for the same {\it design} (i.e., set of stars).  
For example, two plates can have identical targets but be drilled for
observations at different hour angles, temperatures, or epochs, making them unique plates with different plate IDs.
\item[Plate ID] Unique integer assigned to each aluminum plate. 
\item[RJCE] The Rayleigh-Jeans Color Excess method \citep{Majewski_11_RJCE}, a technique for photometrically estimating the line-of-sight reddening to a star,
which APOGEE uses to calculate potential targets' intrinsic NIR colors (\S\ref{sec:colorrange}).
\item[Sky Targets] Empty regions of sky observed on each plate in order to remove the atmospheric 
airglow lines and underlying sky background from the observed spectra (\S\ref{sec:sky}).
\item[Special Targets] Science targets selected with criteria other than the nominal quality, color, and 
magnitude criteria outlined in \S\ref{sec:targselection}.  Examples include known calibration cluster members (\S\ref{sec:calibclusters}) and
ancillary program targets (\S\ref{sec:anc1}).
\item[Targeting Flag \& Bits] A ``flag'' refers to one of two long integers assigned to every target in a design, 
each made up of 31 ``bits'' corresponding to different selection
or assignment criteria (\S\ref{sec:targflags}).  APOGEE's flags are named \atone and {\it apogee\_target2}.  
See Table~\ref{tab:targflags} for a list of the bits currently in use.
\item[Telluric Standards] Blue (hot) stars observed on each plate in order to correct for the telluric absorption lines in the spectra (\S\ref{sec:tellurics}).
\item[Visit] APOGEE's base unit of observation, equivalent to approximately one hour of on-sky integration.  Repeated visits are used both to build up 
signal and to provide an measure of RV stability (e.g., for the detection of stellar companions).
\item[Washington+DDO51] Adopted abbreviation for the combination of Washington $M$ and $T_2$ photometry with {\it DDO51} photometry, 
used in the classification of dwarf/giant stars in many of the halo fields (\S\S\ref{sec:wddata} and \ref{sec:eval-wd}).
\end{description}

\section{Details of Calibration Cluster Target Selection} \label{sec:globclustertargets}
\subsection{Magnitude Distribution and Color Ranges} \label{sec:globclustercmrange}
The complex prioritization scheme and low target density in the halo made
the magnitude sampling of the general survey (\S\ref{sec:magsample}) unnecessary.  Instead, we
selected targets within a priority class starting with the longest cohort of the
field and working our way down to the shorter cohort(s) because ``long'' and
``medium'' cohorts needed to be included in multiple plate designs.  Within each
cohort for a given priority class, the targets were randomly sampled in apparent {\it H}
magnitude.  For some fields, we limited the number of fibers of a given priority
class and cohort type.  This choice increased the total number of targets by
preferentially selecting brighter stars, which require fewer visits.  

The targets with pre-existing high resolution data are sufficiently valuable and rare that they were
selected without a color cut or any other criterion.  Proper motion, radial
velocity, and SEGUE overlap targets were selected with an uncorrected $(J-K_s) \geq 0.3$ color cut.
For the non-cluster members, we extended the intrinsic color limit of
the general survey to $(J-K_s)_0 \geq 0.3$ because the lower average metallicity of
halo stars will tend to make their colors bluer.  The stars without Washington+DDO51 luminosity classifications
were further subdivided into two priority classes based on color: red
($[J-K_s]_0) \geq 0.5$) and blue ($0.3 \geq [J-K_s]_0) < 0.5$).  
As noted in \S\ref{sec:eval-dr}, the use of a dereddened color cut will preferentially exclude very late
type dwarfs due to overcorrected blue colors, improving the giant/dwarf ratio of the sample.

\subsection{Departures from the Cluster Target Selection Algorithm} \label{sec:globclusterexceptions}
In some special cases, the actual target selection deviated from the standard
cluster target selection algorithm.  Here we note those changes and the affected
fields:
\begin{itemize}
\item {\it M3, M53, N4147, N5466, N5634SGR2, VOD1, VOD2, VOD3, 186+42}, and
  {\it 221+84} --- no color cut on proper motion, RV, and SEGUE overlap
  targets.
\item {\it N4147, VOD1, 186+42}, and {\it 221+84} --- the color cuts for the field stars were
  done using uncorrected colors instead of dereddened colors.
\item {\it M3, M53, N5466, N5634SGR2, VOD2}, and {\it VOD3} --- all targets with bit
  $apogee\_target1 = 6$ set (star selected with no dereddening)
  should have bit $apogee\_target1 = 4$ set instead because these targets were
  selected using RJCE-WISE dereddening.
\item {\it N4147} and {\it 186+42} --- Washington+DDO51 stars were selected sequentially in right ascension, but the
  magnitude sampling was still essentially random.
\item {\it VOD1} and {\it 221+84} --- Washington+DDO51 targets were randomized but with unknown seed, so
  they must be reconstructed from the target list.
\item {\it N4147, VOD1, 186+42}, and {\it 221+84} --- the prioritization of targets following
  the selection of giants were red (uncorrected $[J-K_s] \geq 0.5$) dwarfs, then
  blue ($0.3 \leq$ uncorrected $[J-K_s] <$ 0.5) dwarfs, and finally unclassified
  stars with uncorrected $[J-K_s] \geq 0.3$.  This reordering was due to an error
  in the bookkeeping of which stars were unclassified and which were dwarfs.
  While this scheme is suboptimal, these fields will be useful for
  characterizing the effect of the color cuts on dwarfs and unclassified stars.
\item {\it M3, M53, N4147, N5466, N5634SGR2, VOD1, VOD2, VOD3, 186+42}, and {\it 221+84} --- a
  fiber jacket diameter of 70 arcsec was assumed, which was not conservative enough
  to avoid all collisions.  
\item {\it M53} --- The Washington+DDO51 data for {\it M53}, used for the giant/dwarf classifications, come from
  two different observations that overlapped this field.  In comparing the 
  photometry and giant/dwarf classes for stars in both observations, there was a
  great deal of discrepancy.  Most of the differences, however, appear to be due
  to the fact that one of the observations has null values for many of the stars
  in common.  In the cases where both observations have non-null photometry, the
  values agree fairly well ($\lesssim$0.1 mag) in the magnitude range of our
  potential targets.  We were careful to use the photometry from the better
  observation where available and use the photometry from the other observation only if needed.
\item {\it 221+84} --- The open cluster Melotte 111 (Coma star cluster in Coma Berenices at $d = 96$
  pc) is in the field, so Washington+DDO51 dwarfs (Priority 15; \S\ref{sec:calibclusters}) are prioritized over stars without
  Washington+DDO51 photometry (Priority 13--14).
\item {\it M15} --- No Washington+DDO51 photometry was obtained for this relatively low-latitude ($b\sim-27^\circ$) field.
In addition to known cluster members from previous abundance work, proper motion, and RV
studies, probable cluster members were identified based on $(g-i)$ colors within 0.05 mag of the
cluster fiducial \citep[both the fiducial and SDSS $ugriz$ photometry from][]{An_2008_ugrizinGCs}.
\end{itemize}

\section{Details of Ancillary Programs} \label{sec:anc}
\subsection{M31 Globular Clusters} \label{sec:m31gcs}
One ancillary program (PI: ~R.~Schiavon) represents the first use of APOGEE for extragalactic science: the targeting of globular clusters in M31.
By studying the chemical composition and internal kinematics of M31 clusters observed in integrated light 
(i.e., each cluster observed with a single fiber), this program will determine the abundance pattern of M31's old
halo and bulge to an unprecedented level of detail, 
provide insights into the star formation timescales in the halo and bulge,
and constrain the initial mass function of their first stellar generations.
APOGEE will greatly expand upon the set of elemental cluster abundances
obtained in optical studies \citep[e.g.,][]{Colucci_2009_m31globularclusters,Schiavon_2012_m31GCs}
by determining abundances of elements such as O --- the most abundant metal and a key indicator of the timescales for star formation ---
which lack lines at optical wavelengths that can be used in integrated line studies.
Other key elements accessible by APOGEE include C, N, and Na, 
whose abundances based on optical spectra are uncertain or unavailable altogether.
Further, these data will allow for the derivation of internal velocity dispersions for the target sample's massive clusters.

From the initial list of more than 350 M31 globular clusters \citep{Caldwell_2009_m31gcs}, 
$\sim$250 objects brighter than $H = 15.0$ were targeted, along with the M31 core, M32, and M110. 
To isolate the integrated cluster spectra from that of the background (unresolved) M31 stellar populations, 
each observation of a cluster in the vicinity of the M31 bulge
was accompanied by one of a very nearby ``non-cluster'' background region, ideally $\lesssim$10 arcsec offset from the cluster.
As this distance is significantly smaller than the fiber collision radius of APOGEE fibers ($\sim$1.2 arcmin),
simultaneous observations of the cluster and background positions could not be made.  
We adopted a scheme whereby two designs were made, each
containing a mixture of cluster targets and background regions for clusters on the {\it other} design.  
Globular clusters at large M31-centric distances, against a faint stellar background, 
do not require background region counterparts and were instead targeted on both plates.

Both cluster targets and background
regions are considered ``targets'' for this ancillary program and have bit \atone = 18 set.  The background regions are also flagged
as sky fibers on these plates (\attwo = 4), since they serve the same purpose as the regularly-selected sky fibers (\S\ref{sec:sky}) ---
representation of the typical unresolved sky background in the field --- and can be used in the APOGEE data reduction pipeline
in the removal of airglow lines from spectra of both non-cluster and M31 cluster targets.

\subsection{Ages of Red Giants} \label{sec:RGBages}
In addition to the 21 ``APOGEE--{\it Kepler}'' fields (\S\ref{sec:kepler}), 
one ancillary program (PI: C.~Epstein) is targeting two fields observed by the {\it CoRoT} satellite \citep{Baglin_2006_corot}.
These lie on opposite sides of the Galaxy, with {\it COROTA} at $(l,b) \sim (212^\circ, -2^\circ)$ and {\it COROTC} at $(l,b) \sim (38^\circ, -8^\circ)$.
Because {\it CoRoT} stars probe the disk at a range of Galactocentric radii, 
they complement the {\it Kepler} sample stars, most of which lie near the solar circle. 
As with the {\it Kepler} sample, the seismic information available for the {\it CoRoT} stars permits the 
determination of fundamental stellar parameters --- including age ---
and the extension of this type of data to a range of Galactocentric radii 
is invaluable to studies of how Galactic properties evolved over time.

The {\it CoRoT} set of RG stars with seismic detections was selected 
from the sample analyzed in \citet{Mosser_2010_corotseismic}. 
Approximately 120 stars were targeted in the {\it COROTA} field, and approximately 360 in {\it COROTC};
because of the higher number of candidate targets in {\it COROTC}, 
a targeting strategy was employed similar to that for the Galactic Center field {\it GALCEN}: 
division of targets into three one-visit short cohorts and one three-visit medium cohort. 

Both {\it Kepler} and {\it CoRoT} targets in this ancillary program have bit \atone = 22 set, and targets in the {\it Kepler} fields 
also have bit \atone = 27 set.

\subsection{Eclipsing Binaries} \label{sec:EBs}
One ancillary program (PIs: S.~Mahadevan and S.~W.~Fleming)  
is monitoring {\it Kepler} eclipsing binary (EB) systems to derive their dynamical mass ratios. 
Although masses and radii have been measured to the $\sim$3\% level for nearly 300 EBs\footnote{\url{http://www.astro.keele.ac.uk/jkt/debcat/}}
\citep[e.g.,][]{Torres_2010_EBreview}, 
low-mass ($M < 0.8M_{\odot}$) and longer-period ($P > 5$ days) systems remain under-explored.  
The {\it Kepler} dataset (\S\ref{sec:kepler}) is an valuable source of EBs, 
providing nearly continuous, extremely high-precision photometry \citep{Caldwell_2010_keplerinstr} 
that has been used to detect thousands of EBs across a wide range of stellar parameters and orbital periods 
\citep{Coughlin_2011_keplerEBs,Prsa_2011_keplerEBs,Slawson_2011_keplerEBs}.  
When combined with time-series spectroscopy to measure precise RVs \citep[e.g.,][]{Bender_2012_kepler16EB},
these EBs will offer some of the best empirical constraints for next-generation stellar models.  

The sample of {\it Kepler} EBs comes from the catalogs of \citet{Prsa_2011_keplerEBs} and \citet{Slawson_2011_keplerEBs}
and includes targets in two APOGEE fields that overlap {\it Kepler} pointings ({\it N6791} and {\it N6819}).
By design, the target selection imposed a minimum amount of selection cuts in order to explore as 
diverse a range of stellar and orbital properties as possible.  
The sample targets are limited to EBs with $H < 13$ and classified as having a ``detached morphology''
(i.e., excluding those binaries that experience Roche lobe overflow), 
which minimizes the number of model parameters.
In addition to the {\it Kepler} sample, 4 EBs detected using ground-based photometry \citep{Devor_2008_tresEBs},
the well-studied EB system CV~Boo \citep[][]{Torres_2008_CVBoo}, and 
the M dwarf spectroscopic binary GJ~3630 \citep[][]{Shkolnik_2010_specbinaries}
are included as analysis calibrators.

A total of $\sim$115 EB systems in 5 APOGEE fields are targeted, and all have targeting flag bit \atone = 23 set.

\subsection{M Dwarfs} \label{sec:mdwarfs}
M dwarfs make excellent candidates for planet searches due to both their ubiquity and 
the increased RV signal of a planet in the Habitable Zone \citep[HZ;][]{Kasting_1993_HZsinMSstars}, 
relative to the same planet around an F, G, or K star. The M dwarf planet population is beginning to be uncovered, 
with $\sim$30 planetary systems around M dwarfs discovered through RV variations, transits, and microlensing. 
These systems include a possible planet in the HZ around GJ 667C \citep{Anglada-Escude_2012_superearth-GJ667Cmdwarf} 
and the super-Earth GJ 1214b \citep{Charbonneau_2009_superearth-mdwarf}. Due to their intrinsic optical faintness, 
M dwarfs of subtypes later than $\sim$M4 are difficult targets for optical RV and transit searches. 

However, the coming generation of NIR precision-RV planet surveys, such as HPF \citep{Mahadevan_2010_HET-NIRspec} 
and CARMENES \citep{Quirrenbach_2010_CARMENES}, will be able to search efficiently around hundreds of nearby M dwarfs. 
These surveys will require careful target selection in order to sample a range of stellar abundances and slow projected rotational velocities.
APOGEE is particularly well-suited for the study of nearby M dwarfs because 
these stars emit a much higher fraction of their total flux in the NIR $Y$--$H$-band spectral region ($0.9-1.8\mu$m) than in the optical,
enabling the study of later type stars than can be observed with current optical instruments. 

The primary goals of this ancillary program (PIs: S.~Mahadevan, C.~H.~Blake, and R.~Deshpande) are to 
constrain the rotational velocities and compositions of $\gtrsim$1400 M dwarfs and to 
detect their low mass companions through RV variability measurements (Deshpande et al., {\it in prep}).
As $v \sin{i}$ estimates exist in the literature for only $\sim$300 M dwarfs, 
this sample will increase the number of available M dwarf $v \sin{i}$ measurements by nearly a factor of five. 
By using metallicity-sensitive $H$-band features, including some blended K and Ca lines \citep{Terrien_2012_MdwarfFEHcalib}, 
and bootstrapping off targets with previous metallicity estimates, we can derive metallicities for these M dwarfs,
a measurement notoriously difficult to make directly because of their complex spectra. 
Finally, 
the multiplicity of M dwarfs and the rate of both brown dwarf and high-mass giant planet companions to M dwarfs can be probed
via RV variability (along with direct $K$-band imaging; Deshpande et al., {\it in prep}), 
particularly in the subset of M dwarfs that will have $\geq$12 APOGEE epochs,
with time baselines beginning years before dedicated NIR RV planet searches come online.

Targets are drawn from two primary sources: 
the LSPM-North catalog of nearby stars \citep[LSPM-N;][]{Lepine_2005_LSPMNorth} and the 
\citet[][LG11]{Lepine_2011_Mdwarfcatalog} catalog of nearby M dwarfs, which are both proper motion-selected catalogs. 
The LSPM-N sample required a simple color cut of $(V-K)>4$ to select dwarfs of subtype M4 and later; 
the LG11 catalog already includes extensive color and reduced proper motion cuts aimed at selecting M dwarfs. 
For calibration, several targets are included that are known planet hosts, RV standards, 
and/or have previous $v \sin{i}$ or [Fe/H] estimates. 
We also include five M dwarfs that are {\it Kepler} objects of interest \citep{Borucki_2010_keplerintro},
and three L dwarfs \citep{Wilson_2003_CorMASSdwarfs,Phan-Bao_2008_DENISdwarfs}.
In total, $\sim$70\% of APOGEE's
fields contain at least one M dwarf ancillary target (\atone = 19) in all or most of the visits to the field.

\subsection{Members of the Globular Cluster Palomar 1} \label{sec:gc-mems}
Palomar 1 (``Pal 1'' hereafter) is a faint, potentially young globular cluster that may be associated with the
Monoceros Ring or Galactic Anti-center Stellar Stream \citep{Rosenberg_1998_Pal1, Crane_2003_GASSspectra},
and whose spatial position, young age, and extended tidal tails 
\citep{Niederste-Ostholt_2010_Pal1tails}
make it a good candidate for a recently accreted object currently undergoing
disruption by the Milky Way.  
Its metallicity has only recently been estimated from a very small sample of stars, 
whose spectra also suggest other abundance patterns unusual for globular clusters \citep{Sakari_2011_Pal1abundances}.
APOGEE accepted an ancillary proposal (PI: I.~Ivans) to perform the first large-scale
spectroscopic survey of this red, faint, sparse cluster, with the goal of
tightly constraining the cluster's metallicity and exploring its potentially-unusual 
chemistry in even more dimensions.
Since Pal 1 is thought to have $-1.0 < {\rm [Fe/H]} < -0.5$, the
bulk of the APOGEE survey field and cluster stars (e.g., Table~\ref{tab:calibclusters}) will provide a good
comparison sample for a detailed differential analysis,
including a focus on the chemical effects of cluster age, environment, and accretion history.

The cluster targets were selected with a combination of 2MASS and SDSS photometry.
The initial selection comprised 2MASS stars in a 90 arcmin radius near the
cluster center 
$(l,b = 130.065^\circ, 19.028^\circ)$
that satisfied the following criteria:
(i) no neighboring object within 6 arcsec, 
(ii) {\it JHK}$_s$ photometric quality flags of ``A'' and all other contamination flags null,
and 
(iii) $11 < H < 14.5$. 
Then the sample was trimmed to those stars bracketed in $(J-K_s)$ by isochrones at the most probable boundaries
of the cluster age and metallicity: a 12 Gyr isochrone at ${\rm [Fe/H]}=-0.5$ and
a 5 Gyr isochrone at ${\rm [Fe/H]}=-1.0$, assuming $E(J-K_s)=0.112$ and a distance
modulus of 15.76 \citep{Rosenberg_1998_Pal1feh,Rosenberg_1998_Pal1, Sarajedini_2007_ACSglobclusters, Harris_2010_newMWGCs}.

After all the 2MASS-selected candidates were matched with the SDSS DR7 photometric
database \citep{Abazajian_2009_sdssDR7},
an additional cut in $(u, g-r)$ space was performed.  As the few existing
{\it ugr} isochrones for RGB stars are rather inconsistent
at the possible age range of the cluster,
stars along the probable RGB locus were randomly selected, with even sampling between
$0.5< (g-r) <1.6$.  Also included are the three stars identified as cluster members by \citet{Rosenberg_1998_Pal1feh}.
The highest targeting priority was assigned to stars that
satisfied our selection criteria and were within 10 arcmin
of the cluster center.  Targeting priority outside the cluster center was then
assigned randomly.  
In all, $\sim$250 cluster candidate members are being observed as part of
this ancillary program, 
and all have targeting flag bit \atone = 24 set.

\subsection{Bright Optical Calibrators} \label{sec:fabbian}
In addition to the stellar parameter and abundance calibrators targeted in well-studied clusters and the bulge,
the APOGEE disk and halo fields contain a number of field stars that have
been the subject of optical high-resolution spectroscopic studies. 
One ancillary program (PI: D.~Fabbian) is focusing on these targets with a trifold goal:
(i) to compare abundances and atmospheric parameters
derived using APOGEE's IR spectra with those in the literature derived from optical spectra;
(ii) to test advanced 3D stellar model atmospheres 
\citep[e.g.,][]{Fabbian_2010_solarabuncorrections,Fabbian_2012_3Dmodels} as a complement to observations of 
even closer stars with well-determined angular diameters and very accurate parallaxes; 
and
(iii) to establish kinematic and chemical memberships in the Galactic thin disk, thick disk, or halo, which remain uncertain for some 
of the targets.

The target list for this program comprises stars from the IRTF ``Cool Stars'' Spectral 
Library\footnote{\url{http://irtfweb.ifa.hawaii.edu/$\sim$spex/IRTF\_Spectral\_Library/References.html}}
 \citep{Rayner_2009_irtfcoolstars}
and stars observed by \citet{Reddy_2003_FGdwarfs}
and \citet{Reddy_2006_thickdiskabun} at McDonald Observatory, limited to those with $H < 12.5$ falling within
APOGEE's existing disk and halo fields at the time of the first call for ancillary programs.  
Of these objects, roughly half 
are brighter than $H = 5$, the brightest
magnitude permitted for even APOGEE's telluric calibrator standards,
and thus cannot be targeted via the standard mode of observations.
However, these stars can be observed using the alternate observing modes described in \S\ref{sec:magrange}.
All of the $\sim$20 stars have targeting flag bit \atone = 20 set.

\subsection{Massive Stars: Red Supergiants and Their Progenitors} \label{sec:massivestars}
One ancillary project (PI: A.~Herrero) is targeting OB stars and red supergiants (RSGs) in a two-pronged study of obscured massive stars.
The first goal is to compile a spectral library of known OB stars
superior to existing libraries in terms of S/N, which is important for very hot weak-lined stars,
and particularly of OB stars too highly obscured for more traditional UV or optical studies.
These objects' young ages imply that they reflect the present properties of the ISM, and their
rapid evolution means that even a relatively small sample of these stars provides us with multiple snapshots of different evolutionary phases.
As the progenitors of supernovae, OB stars are some of the primary drivers of galactic chemical evolution.
The early-type stars for this part of the ancillary program are drawn from the 
Galactic O-Stars Spectroscopic Survey \citep[][]{Sota_2011_gosss}, 
in which all of the stars have a well-established spectral type and luminosity class.

The second goal of this ancillary program is to observe a number of main-sequence stars and 
RSGs located in the massive star-forming regions near $l \sim 30^\circ$.  
These regions, often referred to collectively as the Scutum Complex \citep[e.g.,][]{Davies_2007_scutumRSGs,Negueruela_2012_stephenson2}, 
contain a large number of massive ($M \lesssim 10^5 M_\odot$)
but highly-obscured clusters in which only the RSGs have been observed spectroscopically.  
Multiple theories to explain this high concentration of massive clusters have been proposed, including 
(i) the interaction of the Milky Way ``long bar'' with the Scutum spiral arm producing a localized starburst
\citep[e.g.,][]{L-C_1999_scutumSFR} and
(ii) the projection along our line of sight of a dense star-formation ring with a Galactocentric radius similar to the length
of the long bar.
The lack of absolute luminosity or robust distance estimates has so far prevented clear discrimination between
these scenarios.  

APOGEE observations will enable higher precision spectroscopic
parallaxes for the RSGs and help establish cluster membership for early-type stars still on the main sequence.
Cluster RSGs for this study are selected to have $8 \leq H \leq 10$, $(J-K) > 0.75$, 
and color index $0.1 < Q_{\rm IR} < 0.4$ \citep[following][]{Comeron_2002_cygOB2},
while the unevolved stars are selected using the IR pseudo-color technique of \citet{Negueruela_2010_westerlund1OBs}
and have magnitudes as faint as $H \lesssim 14.8$.
All of the $\sim$150 targets for this ancillary program have targeting flag bit \atone = 25 set.

\subsection{Kinematics of Young Nebulous Clusters} \label{sec:kinecluster}
Stellar kinematics in star-forming regions are sensitive tracers of the physical processes 
governing the formation and early evolution of stars, planets, and stellar clusters. 
The velocities of young stars can reveal how dynamics within a molecular cloud influence
protostellar mass accretion and the onset of mass segregation and evaporation in stellar clusters 
\citep[e.g.,][]{Tan_2006_starclusterform,Allison_2009_dynmassseg,Cottaar_2012_westerlund1}.
Similarly, kinematically identified protostellar multiple systems are key calibrators 
for pre-main sequence evolutionary tracks \citep[e.g.,][]{Mathieu_1994_PMSbinaries}, 
the influence of age and environment on the binary population \citep[e.g.,][]{Melo_2003_TTaurimultiples,Prato_2007_Ophspecbinaries}, 
and, potentially, the formation mechanisms of planetary systems.

The INfrared Spectroscopy of Young Nebulous Clusters program (IN-SYNC; PIs: K.~Covey and J.~Tan) is conducting
a detailed kinematical survey of the Perseus Molecular Cloud, 
a unique natural laboratory for understanding how gas removal influences the dynamics of young clusters. 
The Perseus cloud is bracketed by two young sub-clusters: IC~348, a $\sim$3 Myr old, 
optically revealed cluster that exhibits evidence of mass segregation 
\citep{Luhman_2003_ic348,Schmeja_2008_embeddedclusters,Muench_2007_ic348}, 
and NGC~1333, a $<$1 Myr old, heavily embedded cluster with little evidence for mass segregation 
\citep{Wilking_2004_ngc1333,Schmeja_2008_embeddedclusters}.
IC~348 and NGC~1333 therefore represent a rare opportunity to compare directly the
kinematical properties of two clusters that share similar initial
conditions but have significantly different present-day evolutionary states.

Since the Perseus cloud is larger than the maximum APOGEE plate FOV, 
distinct fields were defined for each of the two clusters: {\it IC348} and {\it N1333}.
Targets in the IC~348 cluster were selected from catalogs of cluster members
assembled by \citet{Luhman_2003_ic348} and \citet{Muench_2007_ic348}, and from a sample of
candidate cluster members identified at large radii using data from wide field
surveys (e.g., the {\it Spitzer} c2d survey, USNO-B, and 2MASS; Muench et
al., {\it in prep}).  Targets in the NGC~1333 cluster were selected using catalogs
assembled by \citet{Getman_2002_ngc1333}, \citet{Jorgensen_2006_c2dperseus},
\citet{Gutermuth_2008_ngc1333}, and \citet{Winston_2009_ysongc1333serpens,Winston_2010_xrayngc1333serpens},
supplemented with other cluster members identified in numerous studies over
the past two decades (master catalog assembled by Rebull et al., {\it in prep}).

As with the calibration clusters (\S\ref{sec:calibclusters}),
multiple designs were made for each young cluster to resolve fiber
conflicts, in addition to sampling multiple epochs for spectroscopic
binary identification. IC~348 and NGC~1333 were targeted with six and three
distinct designs, respectively.
The highest priority targets in each cluster are those
with $8 < H < 12.5$, 
which were
further sorted according to their extinction-corrected $H_0$ magnitudes (with brighter ones at higher priority) to ensure
that the survey is as complete as possible for higher mass stars.
Once fibers had been allocated to all accessible targets with $H < 12.5$, additional cluster members with $H > 12.5$
were assigned to all designs for that field. 
After accounting for all possible cluster members, any fibers remaining in a design were assigned to normal APOGEE field RG targets.

To mitigate the impact of pre-main sequence (PMS) binary systems on interpretation of the above clusters' velocity dispersions, 
IN-SYNC also targets $\sim$115 PMS members of the cluster NGC~2264 (in both designs of APOGEE field {\it 203+04})
to provide an independent measurement of the frequency of PMS binaries.
Targets in NGC~2264 were selected using catalogs that
identify cluster
members via elevated X-ray emission \citep{Dahm_2007_xrayngc2264}, H$\alpha$
emission \citep{Sung_2008_ngc2264}, or mid-IR excess \citep{Sung_2009_ngc2264}.  Of
these, $\sim$75\% were selected as having $9 < H < 12.9$ and prioritized
for fiber allocation in order of descending $(J-K)$ color. The remaining
$\sim$25\% of the targets were selected from a sample of IR variables identified from
{\it Spitzer} monitoring of NGC~2264 by the YSOVAR program 
\citep{Morales-Calderon_2011_MIR-ONC,Morales-Calderon_2012_PMSEBsONC}
and were included to establish the extent to which IR
variability may limit a PMS star's RV stability.

All targets from the IN-SYNC ancillary program have targeting flag \attwo = 13 set.

\subsection{The Milky Way's Long Bar} \label{sec:longbar}
One outstanding puzzle of the inner MW is the nature of the Galactic ``long bar'',
defined here as the observational feature characterized by increased star counts 
in the near side of the inner disk ($8^\circ \lesssim l \lesssim 30^\circ$), 
to differentiate from the boxy, bar-like bulge and from 
the potential ``nuclear'' bar \citep[][]{Alard_01_centralbar,Nishi_05_centerstruct,Shen_10_purediskbulge,Robin_2012_besanconbarmodel}. 
After it was observed in the early 1990s in mid-IR surface brightness maps \citep{BlitzSpergel_91_GCbar,Dwek_95_DIRBEbar}, 
multiple groups seeking to identify the stellar component of this 
structure found starkly conflicting results on its shape and orientation, 
with a range of line-of-sight angles spanning $\sim$25--60$^\circ$ \citep[e.g.,][]{Hammersley_00_longbar,Benjamin_05_glimpse,L-C_07_longbar}.

Our understanding of the stellar kinematics and chemistry in the bar region is relatively sparse as well.  
We do not know much about how the overall motion of the bar around
the Galactic Center compares to the rotation of the stellar disk or to the rotation of the bulge, nor do we know much of the internal dynamics of the bar,
such as the RV dispersion as a function of galactic longitude or the shapes of stellar orbits trapped in its potential
\citep[though APOGEE has begun to shed light on this latter question; see][]{Nidever_2012_apogeebar}.  
While these parameters are relatively easily obtained for external galaxies and are used to help classify the existence and strength of extragalactic bars
\citep[e.g.,][]{Kuijken_95_peanutbars,Merrifield_99_boxbulgebarrelation,Chung_04_boxybulgekinematics}, 
we have not yet been able to place the MW confidently in sequence with these other bars.
Furthermore, N-body models suggest that bars may be highly effective at migrating stars radially in the inner parts of galaxies, thus
modifying the signatures of mergers and star formation events 
\citep[e.g.,][]{Friedli_94_bargradients,Wozniak_07_barstellarages}.  

One ancillary program (PI: G.~Zasowski) is targeting $\sim$675 long bar RC giants
in 11 fields spanning $8^\circ \leq l \leq 27^\circ$ and $|b| \leq 4^\circ$.
The RC targets for this program were drawn from the 2MASS PSC, {\it Spitzer}-GLIMPSE, and WISE catalogs and selected
as having mid-IR 4.5 $\mu$m magnitudes within 0.4 mag of the mean RC star count peak at each field's $l$, as measured
in GLIMPSE and WISE magnitude distributions, 
which show a clear ``bump'' due to the bar at these longitudes \citep{Zasowski_2012_innerMW}.
In addition, all targets meet 
(i) the same photometric data quality requirements as the normal APOGEE targets (\S\ref{sec:quality}), 
(ii) a dereddened color criteria of $0.5 \leq (J-K_s)_0 \leq 0.8$, and
(iii) a magnitude limit of $H \leq 12.75$ or 13.94, depending on the number of visits planned to the field.

All RC targets from this program have targeting flag bit \attwo = 14 set.

\subsection{Characterization of Early-Type Emission-Line Stars} \label{sec:bestars}
Due to the need to remove telluric absorption from the observed spectra,
APOGEE has targeted and observed nearly three dozen early-type (OBA) stars on each plate since the beginning of the survey;
see the description of these calibrators and their selection in \S\ref{sec:tellurics}.
A small fraction of these stars observed during the first year of APOGEE were found to have
emission-line spectra, dominated by double-peaked Brackett lines,
with only about a third of them noted as emission-line stars in the literature.
Historically, emission-line stars have been identified simply by the presence of such lines in their spectra
\citep[though often with the additional criterion of a NIR or MIR excess; e.g.,][]{Allen_1976_BeIRexcess,Zickgraf_1998_Bedefinition}
and grouped together under the ``Be'' or ``B[e]'' stellar type label, but
the emission is caused by different physical mechanisms that depend on the evolutionary stage of the star.
With the exception of certain stellar types with particularly distinctive spectra (e.g., Wolf-Rayet stars),
most emission-line stars cannot be better characterized without time-intensive, high-resolution spectroscopy.

For many of these objects, the APOGEE spectra provide the first high-resolution, high-S/N view of the emission line profiles.
APOGEE's procedure of visiting (most) fields multiple times enables a multi-epoch analysis of these line profiles,
which can be used not only to identify binary systems but also to track variations in the strengths and profile shapes of the emission lines.
In turn, these variations may be used to 
trace the density structure of circumstellar disks or shells \citep[e.g.,][]{Wisniewski_2007_Be-disks,Stefl_2009_Be-zetaTau-variability}.
In addition to analyzing the sample of serendipitously observed emission-line stars (i.e., those observed as telluric calibrators),
this ancillary program (PI: D.~Chojnowski) is deliberately targeting 25 known Be stars 
\citep[classified via their optical spectra and drawn from the Be Star
Spectra Database;][]{Neiner_2011_Be-database} that fall within APOGEE fields
and have $H < 10$ and $(J-K_s) < 0.5$.

Furthermore, comparison of the Be telluric calibrator subsample with the full APOGEE telluric calibrator sample
will provide 
statistics on the Be/B ratio and population characteristics,
including the enhanced binary fraction of Be stars (compared to single B stars) and 
its connection to the class's spectral properties \citep[e.g,][]{Kriz_1975_Bebinaries,Kogure_2007_emissionlinestars}.
We note that due to the difficulty of identifying luminosity class using $H$-band spectra alone \citep{Steele_2001_Be-Hbandspectra}, 
the APOGEE data will be complemented with optical spectroscopy where possible.

The targets from this program that were selected as known Be stars have targeting flag bit \attwo = 15 set.

\subsection{{\it Kepler} Cool Dwarfs} \label{sec:kepcooldwarfs}
In addition to the extensive {\it Kepler} asteroseismology sample (\S\ref{sec:kepler} and Appendix~\ref{sec:RGBages}), 
APOGEE is targeting $\sim$1200 of {\it Kepler'}s cool dwarfs, observations of which will serve a number of complementary scientific goals
(PI: J.~van~Saders). 
While rotation periods measured from the starspot modulation of {\it Kepler} light curves make it possible 
to extend the sample of field stars with measured ages to include objects 
that are (on average) too faint for asteroseismology,
detailed abundances from APOGEE enable the investigation into metallicity biases in the gyrochronology relations
\citep[e.g.,][]{Mamajek_2008_rotationages,Meibom_2009_m35rotations}, 
which have thus far been neglected. 
This collection of cool dwarfs also represents a valuable comparison sample to the collection of {\it Kepler} planet host candidates,
since only $\sim$3\% of these particular dwarfs have also been identified as potential planet hosts. 
In particular, this sample will facilitate comparisons between the abundances and abundance patterns (e.g., in refractory versus volatile elements) 
of single stars and planet host stars --- relationships
that have important implications for the planet formation process \citep[e.g.,][]{Gonzalez_2006_planethostcomposition}. 

Targets were selected to have 
$7 < H < 11$, $T_{\rm eff} \leq 5500$K, and $\log~g \geq 4.0$, 
where effective temperatures and gravities were obtained from the {\it Kepler} Input Catalog \citep[KIC;][]{Brown_2011_KIC}
with the $T_{\rm eff}$ corrections of \citet{Pinsonneault_2012_KICTeffrevision}. 
An additional $\sim$50 M dwarf candidates that have been continuously monitored by {\it Kepler} with 
magnitudes slightly above the the $H = 11$ faint limit were also included because of their high science impact
--- for example, many stars identified as M dwarfs in the KIC 
have subsequently been classified as giants via gravity-sensitive spectral indices \citep[][]{Mann_2012_coolKeplerstars},
a finding with implications for the interpretation of planet search results and analyses of the frequency and properties of planet hosts.
These late-type stars are drawn from the sample of \citet{Mann_2012_coolKeplerstars}, and while many were already
included in the ancillary target list using the \teff and $\log~g$ requirements described above, 
the remaining (faintest) objects were also included for completeness.

All targets from this ancillary program have targeting flag \attwo = 16 set.

\subsection{Newly Discovered and Unstudied Open Clusters} \label{sec:mirclusters}
While not deviating significantly in essence from the open cluster target selection algorithm and goals described in \S\ref{sec:openclusters},
this ancillary program (PI: R.~L.~Beaton) was granted dedicated fibers because its proposed targets include one cluster discovered by the ancillary team
(Zasowski et al., {\it in press}).
The other clusters were initially identified by the automated cluster search of the 2MASS catalog by 
\citet[][]{Froebrich_2007_clustersearch} --- FSR~0494 and FSR~0665 --- 
but to date, no follow-up study has been made to determine the clusters' basic parameters.
In addition to their selection based on reddening and location in the CMD,
all of the $\sim$13 targets per cluster were constrained to have
photometric uncertainties in $J$, $H$, $K_s$, and $[4.5\mu]$ of $\leq$0.1 mag.

Targets from this program have targeting flag bit \attwo = 17 set.

\end{appendix}

\end{document}